\documentclass[10pt]{article}

\usepackage[preprint]{tmlr}

\usepackage{amssymb}
\usepackage{mathtools}
\usepackage{amsthm}
\usepackage{bm}
\usepackage{graphicx}
\usepackage{subcaption}
\usepackage[space]{grffile}
\usepackage{url}
\usepackage{booktabs}
\usepackage{float}
\usepackage{multirow}
\usepackage{algorithm}
\usepackage{algpseudocode}
\usepackage{tikz}
\usetikzlibrary{arrows.meta,positioning,fit,backgrounds}

\usepackage{hyperref}
\hypersetup{pdftitle={A Feasible Lyapunov Actor Penalty for Soft Actor Critic under Changing Input Limits},pdfauthor={Dhruv S. Kushwaha, Zoleikha A. Biron},colorlinks=true,linkcolor=blue,citecolor=blue,urlcolor=blue}

\graphicspath{{./Images/}}

\newcommand{\ablThreeTrackBackup}{$184.0\pm5.4$}

\newcommand{\ablThreeTrackEuclidDiff}{22.6}
\newcommand{\ablThreeTrackEuclidP}{0.0029}
\newcommand{\ablThreeTrackFixlam}{$192.5\pm3.9$}

\newcommand{\ablThreeTrackFixlamRho}{$120.7\pm11.4$}

\newcommand{\ablThreeTrackHead}{$188.9\pm5.7$}
\newcommand{\ablThreeTrackPrior}{$195.4\pm0.9$}
\newcommand{\ablThreeTrackPriorCI}{[-10.6, -2.3]}
\newcommand{\ablThreeTrackPriorDiff}{-6.4}
\newcommand{\ablThreeTrackPriorHolm}{0.012}

\newcommand{\ablTwoTrackBackup}{$183.8\pm6.9$}

\newcommand{\ablTwoTrackEuclidDiff}{15.7}
\newcommand{\ablTwoTrackEuclidP}{0.0011}
\newcommand{\ablTwoTrackFixlam}{$190.0\pm4.2$}

\newcommand{\ablTwoTrackFixlamRho}{$186.5\pm2.8$}

\newcommand{\ablTwoTrackHead}{$185.5\pm9.8$}
\newcommand{\ablTwoTrackPrior}{$176.3\pm11.1$}
\newcommand{\ablTwoTrackPriorCI}{[-0.7, {+}19.1]}
\newcommand{\ablTwoTrackPriorDiff}{9.2}
\newcommand{\ablTwoTrackPriorHolm}{0.066}

\newcommand{\boxGainSacMax}{107}
\newcommand{\boxGainSacMin}{26}
\newcommand{\boxNpoints}{7}
\newcommand{\boxPSacMax}{1\cdot 10^{-3}}

\newcommand{\boxThreeTrackCLowGainFloor}{-12.0}

\newcommand{\boxThreeTrackCLowHeadFail}{20}
\newcommand{\boxThreeTrackCLowHeadShare}{50}

\newcommand{\boxThreeTrackCLowPrior}{112}
\newcommand{\boxThreeTrackCLowPriorFail}{54}

\newcommand{\boxThreeTrackCLowPriorTFail}{49}

\newcommand{\boxThreeTrackCLowSacFail}{88}

\newcommand{\boxThreeTrackCOneGainFloor}{2.3}

\newcommand{\boxThreeTrackCTwoGainFloor}{13.5}

\newcommand{\boxThreeTrackCTwoHeadShare}{8}

\newcommand{\boxTwoTrackCHalfAbs}{72}

\newcommand{\boxTwoTrackCHalfGainFloor}{4.7}

\newcommand{\boxTwoTrackCHalfHeadFail}{1}
\newcommand{\boxTwoTrackCHalfHeadShare}{80}
\newcommand{\boxTwoTrackCHalfPFloor}{0.063}

\newcommand{\boxTwoTrackCHalfPrior}{127}
\newcommand{\boxTwoTrackCHalfPriorFail}{18}

\newcommand{\boxTwoTrackCHalfPriorTFail}{49}

\newcommand{\boxTwoTrackCHalfSacFail}{82}

\newcommand{\boxTwoTrackCOneGainFloor}{29.0}

\newcommand{\boxTwoTrackCOnePFloor}{1\cdot 10^{-5}}

\newcommand{\boxTwoTrackCTwoGainFloor}{125.3}

\newcommand{\boxTwoTrackCTwoHeadFail}{10}
\newcommand{\boxTwoTrackCTwoHeadShare}{12}
\newcommand{\boxTwoTrackCTwoPFloor}{7\cdot 10^{-11}}

\newcommand{\boxTwoTrackCTwoPrior}{129}
\newcommand{\boxTwoTrackCTwoPriorFail}{46}
\newcommand{\boxTwoTrackCTwoPriorT}{188}

\newcommand{\boxTwoTrackCTwoPriorW}{0.1}
\newcommand{\boxTwoTrackCTwoQp}{165}

\newcommand{\boxTwoTrackCTwoRes}{130}
\newcommand{\boxTwoTrackCTwoResFail}{45}

\newcommand{\boxTwoTrackCTwoSacFail}{4}

\newcommand{\certMinRatioOne}{0.35}

\newcommand{\feasOtherHeadMax}{11}

\newcommand{\feasTwoTrackHeadPostMax}{67}
\newcommand{\feasTwoTrackHeadPostMin}{54}

\newcommand{\feasTwoTrackSacPostMax}{81}
\newcommand{\feasTwoTrackSacPostMin}{77}

\newcommand{\finThreeTrackHead}{189}

\newcommand{\finThreeTrackRes}{165}

\newcommand{\finTwoTrackHead}{191}

\newcommand{\finTwoTrackRes}{165}

\newcommand{\gposTwoTrackAbsMin}{100}
\newcommand{\multTwoTrackAbsMax}{15}

\newcommand{\multTwoTrackHeadMax}{0.14}

\newcommand{\nomThreeStabHead}{206}

\newcommand{\nomThreeStabLqr}{188}

\newcommand{\nomThreeStabRes}{204}

\newcommand{\nomThreeStabSac}{162}

\newcommand{\nomThreeTrackDr}{84}

\newcommand{\nomThreeTrackHead}{189}

\newcommand{\nomThreeTrackLag}{172}

\newcommand{\nomThreeTrackLqr}{179}

\newcommand{\nomThreeTrackRes}{192}

\newcommand{\nomThreeTrackSac}{82}
\newcommand{\nomThreeTrackSacFail}{56}
\newcommand{\nomThreeTrackShape}{110}

\newcommand{\nomTwoStabHead}{190}

\newcommand{\nomTwoStabLqr}{187}

\newcommand{\nomTwoStabRes}{196}

\newcommand{\nomTwoStabSac}{177}

\newcommand{\nomTwoTrackAbs}{166}

\newcommand{\nomTwoTrackBest}{206}

\newcommand{\nomTwoTrackDr}{132}

\newcommand{\nomTwoTrackHead}{186}
\newcommand{\nomTwoTrackHeadFail}{8}
\newcommand{\nomTwoTrackLag}{73}
\newcommand{\nomTwoTrackLagFail}{65}
\newcommand{\nomTwoTrackLqr}{121}
\newcommand{\nomTwoTrackLqrFail}{37}
\newcommand{\nomTwoTrackQp}{163}
\newcommand{\nomTwoTrackQpFail}{21}
\newcommand{\nomTwoTrackRes}{172}
\newcommand{\nomTwoTrackResFail}{12}
\newcommand{\nomTwoTrackResQp}{171}

\newcommand{\nomTwoTrackSac}{132}
\newcommand{\nomTwoTrackSacFail}{19}
\newcommand{\nomTwoTrackShape}{138}

\newcommand{\pThreeStabHeadVsAbs}{0.002}
\newcommand{\pThreeStabHeadVsRes}{0.011}
\newcommand{\pThreeTrackHeadVsAbs}{0.843}
\newcommand{\pThreeTrackHeadVsRes}{0.208}
\newcommand{\pTwoStabHeadVsAbs}{0.003}

\newcommand{\pTwoTrackHeadVsAbs}{0.001}
\newcommand{\pTwoTrackHeadVsRes}{0.017}
\newcommand{\pTwoTrackHeadVsResQp}{0.040}

\newcommand{\piAllHead}{0.99}
\newcommand{\piAllHeadHi}{1.00}
\newcommand{\piAllHeadLo}{0.96}

\newcommand{\retunedThreeTrackCLowCI}{[{+}31.4, {+}52.9]}
\newcommand{\retunedThreeTrackCLowGain}{42.2}
\newcommand{\retunedThreeTrackCLowP}{4\cdot 10^{-6}}
\newcommand{\retunedTwoTrackCHalfCI}{[{+}16.1, {+}24.5]}
\newcommand{\retunedTwoTrackCHalfGain}{20.3}
\newcommand{\retunedTwoTrackCHalfP}{5\cdot 10^{-8}}
\newcommand{\retunedTwoTrackCTwoCI}{[-3.6, {+}17.8]}
\newcommand{\retunedTwoTrackCTwoGain}{7.1}
\newcommand{\retunedTwoTrackCTwoP}{0.18}

\newcommand{\robRfourThreeTrackQpFail}{32}

\newcommand{\robRfourThreeTrackResFail}{24}

\newcommand{\robRfourTwoTrackCvarFiveFail}{100}

\newcommand{\robRfourTwoTrackHeadFail}{64}

\newcommand{\robRfourTwoTrackQpFail}{42}

\newcommand{\robRfourTwoTrackResFail}{43}

\newcommand{\robRfourTwoTrackSacFail}{83}

\newcommand{\robRoneThreeTrackLcFailMax}{21}

\newcommand{\robRoneThreeTrackSacFail}{62}

\newcommand{\robRoneTwoTrackHeadFail}{1}

\newcommand{\robRoneTwoTrackQpFail}{32}

\newcommand{\robRoneTwoTrackResQpFail}{26}

\newcommand{\robRthreeTwoTrackHeadFail}{0}

\newcommand{\robRtwoTwoTrackHeadFail}{0}

\newcommand{\robRtwoTwoTrackResFail}{18}

\newcommand{\tcThreeTrackHead}{190}

\newcommand{\tcTwoTrackHead}{172}
\newcommand{\tcTwoTrackHeadFail}{15}

\newcommand{\worstGapAbs}{75}
\newcommand{\worstGapHead}{7}
\newcommand{\worstGapPrior}{67}

\newcommand{\worstGapRes}{65}

\newtheorem{proposition}{Proposition}

\DeclareMathOperator{\CVaR}{CVaR}

\DeclareMathOperator{\DARE}{DARE}
\newcommand{\Ucal}{\mathcal{U}}

\title{A Feasible Lyapunov Actor Penalty for Soft Actor Critic\\under Changing Input Limits}
\author{\name Dhruv S. Kushwaha \\
      \addr Department of Electrical and Computer Engineering, University of Florida
      \AND
      \name Zoleikha A. Biron \\
      \addr Department of Electrical and Computer Engineering, Clemson University}

\begin{document}

\maketitle

\begin{abstract}
A model of the system can be given to soft actor critic (SAC) as a controller in the loop, as a prior on the actor's actions, or as a penalty on the actor. We study a Lyapunov decrease penalty on the SAC actor, where the Lyapunov function is the Riccati quadratic of a linear model of the tracking error identified from data. First, we make the decrease target attainable at every state: it is the larger of a contraction target and the best value an admissible action can reach, plus a small slack. Second, we show that the penalty is then zero on a sublevel set of the predicted quadratic around the best admissible model action, an ellipsoid in the Riccati metric when that action is interior, with a linear boundary term when it saturates and a radius that shrinks to a slack floor where the plain target cannot be met. Third, we retrain every learner with the input box scaled by 0.35 to 2 on 2D and 3D quadrotor tracking. With one setting, the penalty has a higher mean return than SAC at all \boxNpoints{} pairs of task and box, significant after Holm correction at each, and is within \worstGapHead{} return points of the best compared method at every pair. On 2D tracking, residual SAC, an action prior toward the same regulator and a plain contraction target fall \worstGapRes, \worstGapPrior{} and \worstGapAbs{} points behind the best compared method, respectively, in at least one box each. Even after retuning at every box, the prior trails the penalty by \retunedTwoTrackCHalfGain{} points in the smallest 2D box and \retunedThreeTrackCLowGain{} in the smallest 3D box. On 3D tracking, residual SAC and the penalty do not differ significantly at any tested box. Holding the radius at its slack floor costs up to \boxTwoTrackCTwoGainFloor{} points on 2D tracking. At the nominal box a tuned action prior is better on 3D tracking. All results are in simulation after 24{,}000 training steps, and none is a stability guarantee on the plant.
\end{abstract}

\paragraph{Note on this version:} This version replaces versions 1 to 4 of this preprint and has a new title. We rebuilt the method and repeated all experiments, and several statements of the earlier versions do not hold for the new study. It adds a study in which the input box is scaled and every learner is retrained, an action prior baseline, and ablations of the radius and the metric of the penalty. The Lyapunov function is now computed in tracking error coordinates around a dynamically consistent reference, from a centered model without bias terms, so the spectral normalization and the value anchor of the earlier versions are no longer used. The policy acts around the reference feedforward, which is what makes tracking feasible within the input box. With this change, the constraint improves tracking substantially, instead of trading return for lower variance. The constraint is elastic, not hard, and the potential based shaping baseline is now policy invariant and performs at or above SAC, so the earlier conclusions that a hard Lagrangian constraint is essential and that shaping collapses learning are withdrawn, as is the statement that violations decay monotonically. We also found that a simulator call without an explicit client could change the evaluated quadrotor plant when two simulators share a process. We fixed this issue, so the earlier numbers cannot be compared directly with those reported here.

\section{Introduction}
\label{sec:introduction}

Reinforcement learning (RL) can produce controllers for systems whose models are hard to write down, but on physical systems the time it takes to learn, and what the learned controller does on the way, matter as much as the final return \citep{brunke2022safe}. Quadrotor reference tracking is a good example. The reward is dense, the dynamics are smooth, and yet an off policy learner such as soft actor critic (SAC) \citep{haarnoja2018soft} needs many episodes before it stops crashing, because early policies push the vehicle far from the reference where recovery is hard with limited thrust.

Lyapunov functions give a way to tell the learner which actions move the system toward the reference. Constraining or shaping an RL update with a Lyapunov function is an old idea \citep{perkins2002lyapunov,berkenkamp2017safe,chow2018lyapunov} that has been developed in many directions \citep{han2020actor,dawson2023safe,kushwaha2025review,long2025certifying,zhang2025msacl}. The catch is that a decrease condition is only useful as a constraint if it can be satisfied. With actuator limits and an approximate model, a fixed rate of decrease is often impossible far from the reference, and a constraint that cannot be met either has to be relaxed or adjusted online \citep{chen2025stability} or drives an unregularized dual variable without bound.

A model can also enter the learner as the base of a residual policy or as a prior that pulls the actor toward a controller's action. This paper asks: \emph{when a data driven linear model is given to SAC as a Lyapunov decrease penalty rather than as a controller, what does it buy, and does it survive a change of the input limits that the controller and the learning settings were chosen for?} The second part matters in practice, since actuators are derated, payloads change and safety envelopes are tightened after a controller has been designed. We model these changes by varying the control authority around the feedforward. Answering it required us to settle three parts of the design (Figure~\ref{fig:pipeline}).

\paragraph{The coordinates:} A Lyapunov function for tracking must vanish on the reference. We work in tracking error coordinates relative to a dynamically consistent reference, built from differential flatness for the quadrotors \citep{mellinger2011minimum} and from stable inversion for the cartpole \citep{devasia1996nonlinear}, and we let the policy act around the reference feedforward. An EDMD model \citep{williams2015data,korda2018linear} fitted without bias terms, as in \citet{strasser2026safedmd}, then has zero error as an exact equilibrium, and the discrete algebraic Riccati equation gives a quadratic $W$ with a known contraction rate under the model.

\paragraph{The constraint:} Instead of asking for a fixed decrease, we ask the deterministic action to bring $W$ below the larger of a contraction target and the best value an admissible action can reach, plus a small slack. The best value is the solution of a small box constrained quadratic program, which we solve exactly. The constraint is then feasible at every state, so a sufficiently expressive actor can approximate a strictly feasible selection, and a regularized dual variable, which stays bounded in any case, can return to zero once the constraint is met with a margin. The penalty is then zero on a sublevel set of the predicted quadratic around the best admissible model action, an ellipsoid in the Riccati metric when that action is interior, and its radius cannot shrink when the input box grows.

\paragraph{The evidence:} A model based constraint says nothing directly about the true plant, so we report return and failures on held out episodes, under perturbations the learner never saw, and a model free certificate of the decrease of each frozen policy, scored with the same $W$ for every method. To compare the uses of the model, we retrain every learner with the input box scaled from 0.35 to 2 times its nominal size.

\paragraph{Contributions:}
\begin{enumerate}
    \item A feasible relative decrease penalty for the SAC actor, built in tracking error coordinates in which the model has an exact equilibrium, with an exact box QP for the best admissible decrease, and a characterization of the actions it does not penalize (Sections~\ref{sec:model} and~\ref{sec:constraint}).
    \item A sweep of the input box on both tracking tasks that compares the penalty with residual SAC and an action prior built from the same model, the prior also retuned at every box (Section~\ref{sec:results}).
    \item Ablations of the radius of the zero penalty set, an isotropic replacement of its metric, the threshold and the dual update (Section~\ref{sec:results}).
    \item An evaluation at the nominal box on four quadrotor tasks against SAC, shaped SAC, Lyapunov actor critic, a Lagrangian SAC, domain randomized SAC, residual SAC and fixed regulators, under four unseen perturbations, with a measurement of how often a plain contraction target is unattainable and a model free certificate of the closed loop decrease applied to every method (Sections~\ref{sec:certificate} and~\ref{sec:results}).
\end{enumerate}

\begin{figure}[t]
\centering
\resizebox{\textwidth}{!}{
\begin{tikzpicture}[
    font=\small,
    node distance=7mm,
    box/.style={draw, rounded corners=3pt, thick, align=center, minimum height=12mm, inner sep=4pt, fill=white},
    group/.style={draw, rounded corners=6pt, thick, inner sep=6pt},
    arr/.style={-{Latex[length=2mm]}, thick},
    darr/.style={-{Latex[length=2mm]}, thick, dashed},
    lab/.style={font=\scriptsize, align=center, fill=white, inner sep=1pt}
]
\definecolor{off}{RGB}{65,105,178}
\definecolor{online}{RGB}{47,138,87}
\definecolor{learn}{RGB}{198,111,23}
\definecolor{cons}{RGB}{184,61,90}
\definecolor{cert}{RGB}{101,87,166}

\node[box, draw=off] (collect) {Exciting collection\\12 episodes controller $+$ noise\\12 exploratory episodes};
\node[box, draw=off, right=of collect] (edmd) {EDMD in error coordinates\\$h=\psi(S^{-1}e)$,\ $\psi(0)=0$\\$h^+\approx Ah+Bv$, chosen by LOEO};
\node[box, draw=off, right=of edmd] (dare) {Riccati certificate\\$P=\mathrm{DARE}(A,B,I,\rho I)/s_W$\\$W(h)=h^\top Ph$};
\node[box, draw=off, right=of dare] (gates) {Numerical gates G1 to G5\\residual diagnostic\\$\rho_B/\varepsilon_P\ge 1$ (not a uniform bound)};
\draw[arr, off] (collect) -- (edmd);
\draw[arr, off] (edmd) -- (dare);
\draw[arr, off] (dare) -- (gates);

\node[box, draw=online, below=24mm of collect.south west, anchor=north west] (env) {Plant and reference\\$x_k,\ (\bar x_k,\bar u_k)$\\$e_k=x_k-\bar x_k$};
\node[box, draw=learn, right=9mm of env] (actor) {Actor\\$\pi_\theta(\cdot\mid o_k)$};
\node[box, draw=online, right=9mm of actor] (map) {Residual action map\\$u_k=\bar u_k+c\,D_u a_k$};
\node[box, draw=cons, right=9mm of map] (qp)
    {Best one step value (box QP)\\$W^\ast_k(h)=\min_{v\in\mathcal{U}_k} W(Ah+Bv)$\\[1pt]
     threshold $\tau_k=\max\{(1-\eta)W,\ W^\ast_k+\epsilon_\tau(W+w_0)\}$};
\draw[arr, online] (env) -- node[lab, above=1pt] {$o_k$} (actor);
\draw[arr, learn] (actor) -- node[lab, above=1pt] {$a_k$} (map);
\draw[arr, online] (map.south) -- ++(0,-5mm) -| node[lab, pos=0.27] {$u_k$ executed} (env.south);
\draw[arr, off] (gates.south) |- node[lab, pos=0.75] {$A,\ B,\ P$} ([yshift=7mm]qp.north) -- (qp.north);

\node[box, draw=learn, below=16mm of actor] (aupdate) {Actor update\\SAC loss $+\ \lambda\, g+\tfrac{\rho_a}{2}[g]_+^2$};
\node[box, draw=cons, right=of aupdate] (hinge) {Hinge on the deterministic action\\$\ell=\big(W(Ah+Bv(\tanh\mu_\theta))-\tau\big)_+/(W+w_0)$\\$C=\mathrm{mean}(\ell)$ or $\mathrm{CVaR}_q(\ell)$,\ $g=(C-\zeta)/\zeta$};
\node[box, draw=learn, right=of hinge] (buffer) {Replay buffer\\$(o_k,a_k,r_k,o_{k+1},$\\$e_k,k,\bar u_k)$};
\node[box, draw=cons, below=6mm of hinge] (dual) {Dual: $\lambda\leftarrow\big[(1-\beta\epsilon)\lambda+\beta g\big]_+$};
\draw[arr, learn] ([xshift=-4mm]map.south east) |- node[lab, pos=0.92] {transition} ([yshift=5mm]buffer.north) -- (buffer.north);
\draw[arr, cons] ([xshift=-14mm]qp.south) -- node[lab, right=2pt, pos=0.35] {$W^\ast,\ \tau$} ([xshift=-14mm]qp.south |- hinge.north);
\draw[arr, learn] (buffer) -- (hinge);
\draw[arr, cons] (hinge) -- (aupdate);
\draw[arr, cons] (hinge) -- (dual);
\draw[darr, cons] (dual.west) -| node[lab, pos=0.75] {$\lambda$} (aupdate.south);
\draw[arr, learn] (aupdate) -- node[lab, right=2pt, pos=0.3] {$\theta$} (actor);

\node[box, draw=cert, below=28mm of aupdate.south -| env.west, anchor=north west] (freeze) {Freeze the policy\\selected on validation\\fix $\alpha,\ \omega,\ n=19$};
\node[box, draw=cert, right=of freeze] (run) {Run $n$ fresh episodes\\on the true plant};
\node[box, draw=cert, right=of run] (score) {Episode score\\$s_j=\max_t\,(W(h_{t+1})-\omega)_+/W(h_t)$\\($+\infty$ on failure)};
\node[box, draw=cert, right=12mm of score] (out) {$\hat\rho$ = rank $\lceil(n+1)(1-\alpha)\rceil$ score\\Marginal episode coverage $1-\alpha$\\Useful contraction only if $\hat\rho<1$};
\draw[arr, cert] (freeze) -- (run);
\draw[arr, cert] (run) -- (score);
\draw[arr, cert] (score) -- (out);

\begin{scope}[on background layer]
\node[group, draw=off, fit={(collect)(edmd)(dare)(gates)}, label={[off, font=\bfseries\small, anchor=south west]north west:{1. Identification and certificate, offline (Section~\ref{sec:model})}}] {};
\node[group, draw=online, fit={(env)(qp)(buffer)(aupdate)(dual)([yshift=8mm]qp.north)}, label={[online, font=\bfseries\small, anchor=south west]north west:{2. Learning with the feasible decrease constraint (Section~\ref{sec:constraint})}}] {};
\node[group, draw=cert, fit={(freeze)(out)}, label={[cert, font=\bfseries\small, anchor=south west]north west:{3. Closed loop certification after training (Section~\ref{sec:certificate})}}] {};
\end{scope}
\end{tikzpicture}}
\caption{Overview of LC SAC. (1) An EDMD model of the tracking error is fit on exciting data and gives the Riccati quadratic $W$. (2) During training, the actor loss adds a dual penalty on the mean or CVaR of a relative decrease hinge evaluated at the deterministic action. The threshold uses the best admissible one step value $W^\ast$, which makes the constraint feasible. (3) After training, fresh episodes of the frozen policy give a model free bound $\hat\rho$ on its one step decrease of $W$ on the true plant.}
\label{fig:pipeline}
\end{figure}

\section{Related Work}
\label{sec:related}

\paragraph{Lyapunov methods in RL:} \citet{perkins2002lyapunov} restricted an agent to actions that descend a given Lyapunov function, and \citet{berkenkamp2017safe} certified regions of attraction with a learned model. \citet{chow2018lyapunov,chow2019lyapunov} built Lyapunov functions for constrained Markov decision processes \citep{altman1999constrained}. Lyapunov actor critic (LAC) \citep{han2020actor} learns a Lyapunov critic alongside the policy, and recent work certifies RL policies with generalized \citep{long2025certifying} or multistep \citep{zhang2025msacl} Lyapunov conditions, or gives finite sample probabilistic stability guarantees \citep{han2026reinforcement}. \citet{chen2025stability} combine SAC with a control Lyapunov function and adjust its constraints during learning. Surveys are given by \citet{dawson2023safe}, \citet{gu2024review} and in our review \citep{kushwaha2025review}. Infeasible conditions are a known problem: control Lyapunov function programs relax the decrease with a penalized slack \citep{galloway2015torque}, optimal decay barrier functions stay feasible at every state in the safe set \citep{zeng2021safety}, and statewise constrained RL restricts the constraint to states where it can be met \citep{ma2021feasible,yu2022reachability}. The closest precedent is the contractive constraint of Lyapunov based model predictive control, which asks for the decrease a backup controller achieves \citep{mhaskar2006stabilization,delapena2008lyapunov}. A baseline policy plays this role in \citet{chow2018lyapunov}. Our threshold uses instead the best decrease any admissible action achieves, computed exactly from a data driven Riccati certificate, and applies it to the actor of an off policy learner. At the nominal box we detect no significant return difference between this threshold and the backup rule in the same learner, so what we study is an attainable target in the actor update and how it compares with other uses of the same model. \citet{gill2025off} add a hinge on the decrease of a learned Lyapunov function, with a fixed decay rate, to the SAC actor loss. Our target is attainable by construction and our metric is fixed and identified. Model based controllers also enter RL as the base of a residual policy \citep{silver2018residual,johannink2019residual}, also with a bounded residual and a Lyapunov argument \citep{staessens2021adaptive}, or as a prior that regularizes the actor \citep{cheng2019control}, and Lyapunov functions shape costs \citep{choi2020reinforcement,westenbroek2022lyapunov}. We compare these uses of the same identified model while varying the input box. We retrain each learner separately at every box using the same model and settings.

\paragraph{Koopman and EDMD models for control:} EDMD fits a linear predictor in a space of observables \citep{williams2015data,korda2018linear,brunton2022modern}. With inputs, EDMD extends dynamic mode decomposition with control \citep{proctor2016dynamic}. SafEDMD \citep{strasser2026safedmd} designs controllers with closed loop guarantees from EDMD surrogates whose proportional error bounds vanish at the origin, which requires a model that preserves the origin. Our centered model without bias terms follows the same principle. EDMD models have been used to build barrier and Lyapunov constraints for learning before \citep{folkestad2021data}, including in our earlier work \citep{kushwaha2024lyapunov,kushwaha2026robust}, where we noted that a margin calibrated on data from one collection policy is biased once the learned policy changes the visited states, and our identification therefore uses exciting data.

\paragraph{Risk aware constraints:} The conditional value at risk \citep{rockafellar2000optimization,rockafellar2002conditional} has been used for risk sensitive policy gradients \citep{tamar2015policy}, percentile constraints \citep{chow2018risk} and safety constraints in SAC \citep{yang2021wcsac}. Constrained policy optimization enforces expected costs through Lagrangian or trust region updates \citep{achiam2017constrained,tessler2019reward}. We apply CVaR to the distribution of decrease violations over replay states, where its dual form gives a statement about bounded shifts of that distribution.

\paragraph{Conformal certificates:} Split conformal prediction gives distribution free coverage from exchangeable samples \citep{vovk2005algorithmic,vovk2012conditional,angelopoulos2023conformal}, and it has been used to certify learned Lyapunov and barrier functions \citep{binny2025safe}. We use it on whole episodes of a frozen policy, so that the statement concerns the true plant, not the model.

\section{Problem Setting}
\label{sec:setting}

\paragraph{Plant, reference and error:} The plant is a deterministic simulator $x_{k+1}=F(x_k,u_k)$ with a control period $\Delta t$ and horizon $T$, and the task gives a reference $r_k$ used by the reward. From $r_k$ we compute a dynamically consistent reference $(\bar x_k,\bar u_k)$, $k=0,\dots,T$. For the quadrotors, flatness maps the position reference and its derivatives to attitude and thrust \citep{mellinger2011minimum}. For the cartpole, we compute a periodic stable inversion orbit by collocation \citep{devasia1996nonlinear}. For stabilization, $\bar x_k$ is the goal and $\bar u_k$ the hover or zero input. The tracking error is $e_k=x_k-\bar x_k$ and the dimensionless input deviation is $v_k=D_u^{-1}(u_k-\bar u_k)$ with a diagonal scale $D_u$.

\paragraph{Residual action map:} The policy outputs $a_k\in[-1,1]^m$ and the executed input is $u_k=\bar u_k+D_u a_k$, clipped to the actuator limits. The admissible set of deviations at phase $k$ is the box $\Ucal_k=\{v\in[-1,1]^m: \bar u_k+D_u v \text{ within limits}\}$. On the 2D circle the feedforward lies outside the box that an absolute action map would allow on 57\% of phases, so the residual map is what makes tracking feasible. We use the residual map for every method, including SAC and the baselines. In the box sweep of Section~\ref{sec:results} the map is $u_k=\bar u_k+c\,D_u a_k$, so $v=c\,a$ and $\Ucal_k$ scales with $c$.

\paragraph{Learner:} The actor observes $o_k=[x_k, e_k, k/T]$, and on tracking tasks also $r_{k+1}$. The reward is the benchmark's $\exp(-c_k)$ with a quadratic cost $c_k$ in the state error and the input, so each step earns at most one and a return of $T$ is a perfect episode. Episodes end at $T$ or when the state leaves its bounds, which we count as a failure. For the quadrotors, the bounds include $|x|,|y|\le 2$\,m, $z\le 2$\,m and roll and pitch below 85 degrees, and the top of the 2D circle lies on $z=2$\,m.

\section{Model and Lyapunov Certificate}
\label{sec:model}

\paragraph{Identification data:} For each quadrotor task we record 24 episodes from initial states inside the state bounds with $z\ge 0.4$\,m (and $|x|\le 0.4$\,m on 2D tracking): 12 with a discrete LQR, designed on the nominal model linearized at hover with diagonal weights and acting on the tracking error, plus Gaussian input noise of standard deviation 0.15, and 12 exploratory episodes with the same LQR at gain 0.5 (0.8 on 2D tracking) plus uniform input noise of amplitude 0.6 (0.4), all clipped to the input box. On the cartpole, 12 episodes use the benchmark's LQR with Gaussian input noise and 12 use inputs drawn uniformly from the box. Rows after the first ground contact are removed, since post crash data lie outside the identification domain. These data give much better models than an earlier quadrotor collection with the same number of episodes (Table~\ref{tab:certificate} and Figure~\ref{fig:certificate} in the appendices), which used the benchmark's PID with input noise and uniform inputs from unrestricted initial states and whose quadrotor episodes lasted 20 to 43 steps on average: its worst fold margin ratio, defined below, is as low as \certMinRatioOne, against at least 3.60 on the quadrotors with the final data (the cartpole data are the same in both). Within the final collection, the 12 controller episodes alone already give at least 3.19 on the quadrotors.

\paragraph{Model class and selection:} We scale the error as $y=S^{-1}e$ and consider lifts $h=\psi(y)=[y;\ \phi(y)-\phi(0)]$ with thin plate features $\phi$, so that $\psi(0)=0$, $h=\psi(S^{-1}e)$, and the identity block is kept. The predictors are $h^+\approx A h+Bv$ (linear), a bilinear extension with terms $v\otimes h$, a phase scheduled version for tracking, and the same predictors on the thin plate lifts. None of them has a constant column, so $h=0, v=0$ is an exact equilibrium of every candidate. The family and ridge parameter are chosen by leave one episode out cross validation on the identity channels. On the cartpole the search covers all candidates and picks the linear model with $\psi$ equal to the identity on tracking and the bilinear model on stabilization. On the quadrotors we restricted the search to the linear model with $\psi$ equal to the identity and chose only its ridge parameter. Among the other candidates, the thin plate lift on 2D and the bilinear model on 3D have a lower cross validation error and also pass the gates G1 to G5 below, and we did not train with them. We therefore make no claim about nonlinear observables in this paper.

\paragraph{Riccati certificate:} With the linear part $(A,B)$ of the selected model we solve $P_0=\DARE(A,B,I,\rho I)$, where $\rho\in\{1,0.1,0.01\}$ maximizes the guaranteed contraction rate $\gamma=\lambda_{\min}\big(P_0^{-1/2}(I+K^\top\rho K)P_0^{-1/2}\big)$ of the unsaturated regulator $v=-Kh$ \citep{anderson1990optimal}. We set $W(h)=h^\top P h$ with $P=P_0/s_W$, where $s_W$ makes the mean of $W$ on the training rows equal to one. For the linear predictor used in the quadrotor experiments, wherever $-Kh$ lies in the box, $W(Ah-BKh)\le(1-\gamma)W(h)$.

\paragraph{Gates:} A model is admissible only if the Riccati solution passes numerical residual checks and $A-BK$ is Schur (G1), $P_0\succeq I$ (G2), $\mathrm{cond}(P_0)\le 10^6$ (G3), $\gamma\ge 10^{-3}$ (G4), and $B$ has full column rank (G5), which makes the quadratic program for $W^\ast$ below strictly convex with a unique minimizer. We also compare the authority of the box, $\rho_B=\max_{v}\|Bv\|_P$ over its vertices, with the root mean square one step model residual $\varepsilon_P=\|P^{1/2}(h^+-\hat h^+)\|$ on each held out episode, and call a task certificate supported when $\rho_B/\varepsilon_P\ge 1$ on every leave one episode out fold, a diagnostic and not a uniform bound on the model error. All six tasks are certificate supported with the final data. The resulting $\gamma$ lies between 0.0125 and 0.036 per step, the closed loop spectral radius between 0.926 and 0.989, and $\mathrm{cond}(P)$ is at most 221 (Table~\ref{tab:certificate}, appendix).

\section{The Feasible Relative Decrease Constraint}
\label{sec:constraint}

\paragraph{Definition:} For a replay sample with lifted error $h$ and phase $k$, let $\hat h^+(v)=Ah+Bv$ and
\begin{equation}
W^\ast_k(h)=\min_{v\in\Ucal_k} W(\hat h^+(v)),\qquad
\tau_k(h)=\max\big\{(1-\eta)W(h),\ W^\ast_k(h)+\epsilon_\tau\,(W(h)+w_0)\big\},
\label{eq:threshold}
\end{equation}
with $\eta=\gamma/2$, $\epsilon_\tau=10^{-3}$ and $w_0=10^{-2}$. Given the deterministic actor action $a=\tanh\mu_\theta(o)$ and its deviation $v(a)\in\Ucal_k$, the normalized hinge is
\begin{equation}
\ell(o;\theta)=\frac{\big(W(\hat h^+(v(a)))-\tau_k(h)\big)_+}{W(h)+w_0}.
\label{eq:hinge}
\end{equation}
Where a contraction by the factor $1-\eta$ is possible up to the slack, the constraint asks for it. Elsewhere, it asks for the best achievable value up to a small slack. Since $W(\hat h^+(v))$ is a strictly convex quadratic in $v$ when $B^\top PB\succ 0$, $W^\ast_k$ is a box QP in $m\le 4$ variables, which we solve exactly by enumerating the $3^m$ active sets in a batch. $W^\ast$ and $\tau$ depend only on the replay sample, so they carry no gradient. Only $a$ contributes to the actor gradient.

\begin{proposition}[Feasibility]
\label{prop:feasible}
Assume $P\succ 0$, $\Ucal_k$ is a nonempty box containing the origin in its interior, and $B^\top PB\succ 0$. (a) For every $h$ and $k$ the minimizer $v^\ast_k(h)$ of $W^\ast_k$ exists, is unique and satisfies $\ell=0$. (b) If $\epsilon_\tau>0$ and $w_0>0$, a shrunken minimizer $(1-\bar s)v^\ast_k(h)$ with $\bar s>0$ independent of $h$ on a compact set also satisfies $\ell=0$ and lies in the interior of $\Ucal_k$. (c) Consequently, for every tolerance $\zeta>0$ and every compact set of observations, there is a ReLU network actor of sufficient width whose hinge is below $\zeta$ everywhere, so both the mean and the CVaR of $\ell$ under any distribution on that set can be made strictly smaller than any tolerance, so the constraint $C(\theta)\le\zeta$ is strictly feasible.
\end{proposition}

Part (a) holds because $\tau_k(h)\ge W^\ast_k(h)$. Part (b) uses the slack $\epsilon_\tau(W+w_0)$, and part (c) follows from (b) and universal approximation of the continuous selection. The proof is in Appendix~\ref{app:proofs}. The contraction branch is active when $(1-\eta)W(h)\ge W^\ast_k(h)+\epsilon_\tau(W(h)+w_0)$. Otherwise, the threshold is the best admissible model value plus slack. That value may be above $W(h)$, so feasibility does not mean that $W$ decreases, on the model or on the plant. The argument uses only that the reference action is admissible, continuous in $h$ and meets the threshold, so it also holds when $W^\ast_k(h)$ in \eqref{eq:threshold} is replaced by $W(\hat h^+(v_b))$ for a backup action $v_b=\mathrm{clip}(-Kh,\Ucal_k)$, the rule of Lyapunov based MPC. We compare the two thresholds in Section~\ref{sec:results}.

\paragraph{The geometry of accepted actions:} The next result describes the actions with zero penalty and how they depend on the box.

\begin{proposition}[Zero penalty set]
\label{prop:deadzone}
Under the assumptions of Proposition~\ref{prop:feasible}, write $q(v)=W(Ah+Bv)$, $G=B^\top PB$, $t=(1-\eta)W(h)$, and $\delta=\epsilon_\tau(W(h)+w_0)$. For every admissible $v$,
\begin{equation}
q(v)-W^\ast_k(h)=\|v-v^\ast_k\|_G^2+\nabla q(v^\ast_k)^\top(v-v^\ast_k),
\label{eq:gap}
\end{equation}
where the linear term is nonnegative. With the radius $r_k=\max\{t-W^\ast_k,\delta\}$, the zero penalty set is
\begin{equation}
\mathcal Z_k=\left\{v\in\Ucal_k:
\|v-v^\ast_k\|_G^2+\nabla q(v^\ast_k)^\top(v-v^\ast_k)\le r_k\right\}.
\end{equation}
It is contained in the ellipsoid $\|v-v^\ast_k\|_G^2\le r_k$. If $v^\ast_k$ is interior, it equals that ellipsoid intersected with $\Ucal_k$. Enlarging the box (larger $c$) cannot increase $W^\ast_k$, so the radius $r_k$ cannot decrease. When $t\ge W^\ast_k+\delta$, $\mathcal Z_k=\{v\in\Ucal_k:q(v)\le t\}$. Otherwise, the radius is at its slack floor, $r_k=\delta$. In policy units $a=v/c$, the metric is $c^2G$.
\end{proposition}

The proof is a Taylor expansion of the quadratic $q$ and the optimality condition of $v^\ast_k$ (Appendix~\ref{app:proofs}). The penalty therefore asks for any action whose predicted decrease is good enough, and how much freedom it leaves depends on the box. A controller anchored prior, in contrast, adds $w\|a-v_b/c\|^2=(w/c^2)\|v-v_b\|^2$ at every state, so its strength in input units changes with $c$ unless $w$ is retuned, and it pulls toward $v_b$ also where $v_b$ is not the best admissible action. Section~\ref{sec:results} tests both consequences.

\paragraph{Multistep version:} With a horizon $H>1$, we replace the one step prediction by the $H$ step prediction $\hat h_H=A^H h+\sum_{j<H}A^{H-1-j}Bv_j$, the contraction factor $1-\eta$ by $(1-\eta)^H$, and $W^\ast$ by the best $H$ step value over $v_0\in\Ucal_k$ with the continuation $v_{1:H-1}$ fixed to an approximate minimizer computed by accelerated projected gradient. The first step problem is again an exact box QP, strictly convex when $A^{H-1}B$ has full column rank, so the hinge is zero at its minimizer and Proposition~\ref{prop:feasible} carries over. Multistep decrease conditions are not new \citep{long2025certifying,zhang2025msacl}. We include $H\in\{1,5,10\}$ as an ablation.

\paragraph{Aggregation:} The constraint is $C(\theta)\le\zeta$ with $C$ either the batch mean of $\ell$ or its conditional value at risk at level $q=0.75$,
\begin{equation}
\CVaR_q(\ell)=\min_{s}\Big\{s+\tfrac{1}{1-q}\,\mathbb{E}(\ell-s)_+\Big\}.
\end{equation}

\begin{proposition}[Mean and CVaR constraints]
\label{prop:cvar}
Let $\ell\ge 0$ have distribution $\mu$ over replay states. (a) $\{\mathbb{E}_\mu\ell\le(1-q)\zeta\}\subseteq\{\CVaR^\mu_q(\ell)\le\zeta\}\subseteq\{\mathbb{E}_\mu\ell\le\zeta\}$. (b) For any state distribution $\nu$ with density $d\nu/d\mu\le 1/(1-q)$, $\mathbb{E}_\nu\ell\le\CVaR^\mu_q(\ell)$. So the CVaR constraint on replay bounds the mean violation under every on policy distribution whose density with respect to replay is at most four at $q=0.75$, while a mean constraint $\mathbb{E}_\mu\ell\le\zeta$ only gives $\mathbb{E}_\nu\ell\le\zeta/(1-q)$ under the same shift.
\end{proposition}

Both parts follow from the dual representation of CVaR \citep{rockafellar2002conditional} (Appendix~\ref{app:proofs}). Part (a) orders the two constraints at a common tolerance, and no single tolerance makes them comparable for every distribution of violations. We therefore match their \emph{initial relative tightness}: after warm up, we compute the hinge $\ell_{\mathrm{ref}}$ of the feedforward only action $a\equiv 0$ on the shared warm up buffer and set $\zeta=c_\zeta\, C(\ell_{\mathrm{ref}})$ with $c_\zeta=0.1$ for each aggregate. The feedforward action then has the same ratio of aggregate to tolerance under both constraints, although the initial actor need not. Part (b) identifies the distribution shifts covered by the CVaR constraint. Whether its density ratio bound holds during training remains unknown because we do not observe that ratio.

\paragraph{Adaptive quantile:} We also test a heuristic schedule that raises $q$ when the policy moves away from the replay data: $q_t=\mathrm{clip}(1-1/\hat B_t, 0.5, 0.9)$ with $\hat B_t=2+\delta_t/0.25^2$, where $\delta_t$ is the mean squared difference between the current deterministic action and the replay action on the batch. This is motivated by part (b) but does not estimate the state density ratio, and in practice, on the quadrotors, it stays between 0.83 and 0.90 on every update, with a mean between 0.88 and 0.90 per run (Figure~\ref{fig:q} in the appendices).

\paragraph{Actor and dual updates:} With $g=(C-\zeta)/\zeta$, the actor minimizes the SAC objective plus $\lambda g+\frac{\rho_a}{2}[g]_+^2$, and after each actor step the dual variable is updated by
\begin{equation}
\lambda\leftarrow\big[(1-\beta\epsilon)\lambda+\beta g\big]_+,\qquad \beta=10^{-3},\ \epsilon=0.1,\ \rho_a=1.
\end{equation}
Starting from $\lambda=0$, the regularization $\epsilon$ keeps $\lambda\le\sup_n [g_n]_+/\epsilon$ without a cap and returns $\lambda$ to zero in finitely many updates whenever $g$ stays below a negative constant. At a fixed point the constraint is elastic, with $\lambda=[g]_+/\epsilon$, so it acts as a quadratic penalty and not as a hard constraint. We make no convergence claim for the coupled actor, critic and dual dynamics (Appendix~\ref{app:dual}). The penalty is evaluated at the deterministic action, which is the action that is deployed and certified. With a sampled action, $\mathbb{E}\,W(\hat h^+)$ would contain the term $\mathrm{tr}(B^\top PB\,\mathrm{Cov}\,v)$ and penalize exploration noise. Algorithm~\ref{alg:lcsac} in the appendices summarizes the update.

\section{A Model Free Certificate of the Trained Policy}
\label{sec:certificate}

The constraint is imposed on model predictions over replay states, so it does not imply that the trained policy decreases $W$ on the plant. To measure that directly, we freeze the validation selected deterministic policy, run $n=19$ fresh episodes from the task's initial distribution with seeds disjoint from training and validation, and score each episode by
\begin{equation}
s_j=\max_{0\le t<T_j}\frac{\big(W(h_{t+1})-\omega\big)_+}{W(h_t)},\qquad s_j=+\infty \text{ if episode } j \text{ fails},
\end{equation}
with $\omega=0.05$, $0/0=0$ and a positive value over zero set to $+\infty$. Let $\hat\rho$ be the $\lceil(n+1)(1-\alpha)\rceil$ smallest score, which for $n=19$ and $\alpha=0.05$ is the largest one.

\begin{proposition}[Closed loop certificate]
\label{prop:conformal}
If the plant is deterministic, the policy is fixed before the $n+1$ episodes and their initial states are i.i.d., then with probability at least $1-\alpha$ the score of a new episode is at most $\hat\rho$. On this event, if $\hat\rho<\infty$, the new episode does not fail and satisfies $W(h_{t+1})\le\hat\rho\,W(h_t)+\omega$ for all $t<T$. On the same event, if $\hat\rho<1$, then also $W(h_t)\le\hat\rho^{\,t}W(h_0)+\omega/(1-\hat\rho)$.
\end{proposition}

The statement follows from exchangeability of the episode scores \citep{vovk2005algorithmic,angelopoulos2023conformal}. The probability is over the $n$ certification episodes and the new one together, with initial states from the benchmark's distribution restricted to the state bounds. This marginal probability does not condition on the observed $\hat\rho$. The bound holds regardless of model accuracy because the model enters only through the choice of $W$. Because the same $W$ can score any controller, we apply it to the learners and to the clipped EDMD regulator. A value $\hat\rho<1$ certifies practical contraction over the horizon for the task's initial distribution. A value above one, or infinite because an episode failed, gives no such statement.

\section{Experimental Setup}
\label{sec:setup}

\paragraph{Tasks:} We use the 2D and 3D quadrotors and the cartpole of Safe Control Gym \citep{yuan2022safe}, whose quadrotors run on PyBullet \citep{panerati2021learning}.\footnote{Code will be released upon acceptance.} Each system has a stabilization task and a tracking task: a circle of radius 1\,m in the vertical plane for the 2D quadrotor, an ellipse with semi axes 1 and 0.71\,m in a plane tilted by 45 degrees for the 3D quadrotor, and a sinusoidal cart motion for the cartpole, each with a period of five seconds. The quadrotors run at 50\,Hz for $T=250$ steps with the policy acting in the benchmark's normalized action range, $\pm 10\%$ of hover thrust, here centered on the feedforward (the hardware thrust limits never bind). The cartpole runs at 15\,Hz for $T=150$ steps. Initial states are drawn by the benchmark.

\paragraph{Methods:} The SAC based learners share the SAC implementation \citep{haarnoja2018soft,haarnoja2018applications} with twin critics \citep{fujimoto2018addressing}, two hidden layers of 128 ReLU units, Adam \citep{kingma2015adam} with learning rate $3\cdot 10^{-4}$, batch 256, discount factor 0.99 and a fixed entropy temperature of 0.2. Table~\ref{tab:settings} in the appendices lists every setting. We compare each method with SAC. Shaped SAC adds the potential based term $0.5\,(W(h_k)-0.99\,W(h_{k+1}))$ to the reward. This shaping preserves optimal policies \citep{ng1999policy}. Lyapunov actor critic (LAC) uses its own architecture and quadratic cost \citep{han2020actor}. Lagrangian SAC learns a cost critic for the same hinge evaluated at the executed action and enforces its discounted average below the matched tolerance $\zeta$ with a PI multiplier (gains 0.1 and $10^{-3}$) \citep{stooke2020responsive}. SAC with domain randomization \citep{tobin2017domain,peng2018sim} redraws mass, inertia and wind scales in every training episode from $[0.97,1.03]$, $[0.90,1.10]$ and $[0,0.5]$, respectively. Each test perturbation falls outside this training box on at least one axis.

Residual SAC \citep{silver2018residual,johannink2019residual} takes the action $\mathrm{clip}(v_K/c+a,-1,1)$ with $v_K$ the clipped EDMD regulator below in input deviation units, $c=1$ at the nominal box, and $a$ the SAC output, initialized at zero. On 2D tracking we also test the saturation aware regulator as the base. The action prior adds $w\|a-v_b/c\|^2$ to the SAC actor loss at the deterministic action, with $v_b=\mathrm{clip}(-Kh,\Ucal_k)$ \citep{cheng2019control}. We select $w$ from $\{0.1,1,10,100\}$ using 2D tracking at $c=1$ on training seeds 10000, 20000 and 30000. We maximize the mean over seeds of the best checkpoint validation return, breaking ties in favor of smaller $w$, and use the selected weight for both tracking tasks.

For LC SAC we compare mean and CVaR aggregation, test horizons $H\in\{1,5,10\}$, and evaluate the adaptive quantile. The threshold ablations use the plain threshold $\tau=(1-\eta)W(h)$ without the feasibility branch or the backup threshold with the clipped EDMD regulator as $v_b$. We also test a fixed multiplier equal to the median dual value during training of the headline variant (0.107 on 2D tracking and 0.0073 on 3D tracking, each the median over ten seeds of the per run medians), with and without the quadratic term. The remaining ablations use an isotropic distance surrogate $W^\ast_k+\mu\|v-v^\ast_k\|^2$ with $\mu$ the mean eigenvalue of $G$ (which also omits the boundary term in Eq.~\eqref{eq:gap}), or fix the radius at its slack floor, $\tau_k=W^\ast_k+\epsilon_\tau(W+w_0)$. We also evaluate three fixed controllers: the benchmark's own controller (PID for the quadrotors, LQR for the cartpole), clipped to the same box, the box clipped EDMD regulator $v=\mathrm{clip}(-Kh,\Ucal_k)$, and a saturation aware EDMD regulator that minimizes the one step cost $\rho\|v\|^2+(Ah+Bv)^\top P_0(Ah+Bv)$ over $\Ucal_k$, which equals $-Kh$ wherever that action is admissible, with the same exact box QP as $W^\ast$.

\paragraph{Protocol:} Each learner trains for 24{,}000 steps (the length of 96 full quadrotor episodes), of which the first 3{,}000 are warm up. The baselines, the plain threshold ablation and the headline variant, LC SAC with mean aggregation and $H=1$, use ten seeds. The action prior and the ablations at the nominal box use ten on 2D tracking and five on 3D tracking. The other LC SAC variants and the cartpole use five. Every 3{,}000 steps the deterministic policy is scored on six validation episodes, and the best checkpoint is evaluated on twelve test episodes and on the 19 certification episodes of Section~\ref{sec:certificate}. All seed sets are fixed, disjoint, and chosen only so that the initial state is inside the state bounds. The perturbation study evaluates the same checkpoints without retraining under (R1) mass $+5\%$, which uses half of the upward range of the input box, and inertia $+20\%$; (R2) a persistent crosswind and downdraft with sinusoidal gusts; (R3) mass $+4\%$, inertia $+15\%$ and 75\% of the R2 wind; and (R4) mass $-5\%$. R1 to R3 push the vehicle down and R4 pushes it up.

\paragraph{Statistics:} We report the mean and standard deviation over seeds of the per seed mean test return. A dagger marks methods that beat SAC under a Welch $t$ test \citep{welch1947generalization} with Holm correction \citep{holm1979simple} over the arms of the same task at level 0.05. For the headline variant we also report the probability of improvement $P(X>Y)$ over seeds \citep{agarwal2021deep}, averaged over the four quadrotor tasks, with a 95\% stratified bootstrap interval. We read differences from SAC that do not pass the corrected test as inconclusive. Comparisons of the headline variant with one other arm report the uncorrected Welch $p$ value, except that the action prior is compared at the nominal box with Holm correction over the two tracking tasks, the retuned prior at three boxes with correction over the three, and the box sweep with correction over $c$ within each task. We give 95\% Welch intervals for these differences and do not read a nonsignificant difference as equivalence.

\section{Results}
\label{sec:results}

\begin{table}[t]
\centering
\caption{Test return (mean $\pm$ SD over seeds, maximum $T$) of the validation selected policy after 24{,}000 steps. We use ten seeds for the baselines, the headline variant ($H{=}1$, mean) and the plain threshold on the quadrotors, five for the other rows and on the cartpole. Bold: best learner per task. $^\dagger$: better than SAC after Holm correction. In brackets: failure rate in \% of test episodes. The fixed controllers are scored on the 19 certification episodes. The benchmark PID clipped to the same box fails every quadrotor episode. An n/a entry marks a run we did not perform. The QP base differs materially from the clipped regulator only on 2D tracking.}
\label{tab:nominal}
\resizebox{\textwidth}{!}{
\begin{tabular}{lrrrrrr}
\toprule
Method & 2D stab. & 2D track. & 3D stab. & 3D track. & Cartpole stab. & Cartpole track. \\
\midrule
SAC & 177\,$\pm$\,11 [0] & 132\,$\pm$\,8 [19] & 162\,$\pm$\,19 [3] & 82\,$\pm$\,20 [56] & 114\,$\pm$\,1 [8] & 112\,$\pm$\,1 [17] \\
Shaped SAC & 179\,$\pm$\,8 [0] & 138\,$\pm$\,9 [12] & 168\,$\pm$\,20 [0] & 110\,$\pm$\,25$^\dagger$ [18] & 45\,$\pm$\,32 [27] & 109\,$\pm$\,4 [17] \\
LAC & 63\,$\pm$\,33 [47] & 20\,$\pm$\,6 [98] & 16\,$\pm$\,3 [100] & 8\,$\pm$\,2 [100] & 19\,$\pm$\,12 [97] & 5\,$\pm$\,0 [100] \\
Lagrangian SAC & 184\,$\pm$\,4 [0] & 73\,$\pm$\,29 [65] & 190\,$\pm$\,10$^\dagger$ [0] & 172\,$\pm$\,19$^\dagger$ [1] & n/a & n/a \\
Domain rand. SAC & 179\,$\pm$\,7 [0] & 132\,$\pm$\,7 [19] & 178\,$\pm$\,12 [0] & 84\,$\pm$\,24 [53] & n/a & n/a \\
Residual SAC & \textbf{196\,$\pm$\,1}$^\dagger$ [0] & 172\,$\pm$\,12$^\dagger$ [12] & 204\,$\pm$\,2$^\dagger$ [0] & \textbf{192\,$\pm$\,2}$^\dagger$ [0] & n/a & n/a \\
Residual SAC, QP base & n/a & 171\,$\pm$\,18$^\dagger$ [19] & n/a & n/a & n/a & n/a \\
\midrule
LC SAC, mean, $H{=}1$ & 190\,$\pm$\,3 [0] & 186\,$\pm$\,10$^\dagger$ [8] & 206\,$\pm$\,2$^\dagger$ [0] & 189\,$\pm$\,6$^\dagger$ [0] & 114\,$\pm$\,0 [8] & 112\,$\pm$\,0 [17] \\
LC SAC, plain threshold & 185\,$\pm$\,4 [0] & 166\,$\pm$\,12$^\dagger$ [12] & 196\,$\pm$\,8$^\dagger$ [0] & 188\,$\pm$\,8$^\dagger$ [0] & n/a & n/a \\
LC SAC, mean, $H{=}5$ & 183\,$\pm$\,3 [0] & 201\,$\pm$\,7$^\dagger$ [2] & 195\,$\pm$\,6$^\dagger$ [0] & 148\,$\pm$\,19$^\dagger$ [5] & \textbf{114\,$\pm$\,0} [8] & 113\,$\pm$\,0 [17] \\
LC SAC, CVaR, $H{=}1$ & 188\,$\pm$\,3 [0] & 184\,$\pm$\,3$^\dagger$ [8] & \textbf{206\,$\pm$\,2}$^\dagger$ [0] & 186\,$\pm$\,9$^\dagger$ [0] & 114\,$\pm$\,0 [8] & 112\,$\pm$\,1 [17] \\
LC SAC, CVaR, $H{=}5$ & 185\,$\pm$\,6 [0] & \textbf{206\,$\pm$\,2}$^\dagger$ [0] & 200\,$\pm$\,4$^\dagger$ [0] & 169\,$\pm$\,14$^\dagger$ [0] & 112\,$\pm$\,4 [10] & \textbf{113\,$\pm$\,0} [17] \\
LC SAC, CVaR, $H{=}10$ & 180\,$\pm$\,6 [0] & 202\,$\pm$\,1$^\dagger$ [0] & 194\,$\pm$\,4$^\dagger$ [0] & 162\,$\pm$\,12$^\dagger$ [0] & 114\,$\pm$\,1 [8] & 112\,$\pm$\,0 [17] \\
LC SAC, adaptive $q$, $H{=}1$ & 187\,$\pm$\,3 [0] & 186\,$\pm$\,5$^\dagger$ [8] & 206\,$\pm$\,2$^\dagger$ [0] & 190\,$\pm$\,8$^\dagger$ [0] & 114\,$\pm$\,0 [8] & 112\,$\pm$\,0 [17] \\
LC SAC, adaptive $q$, $H{=}5$ & 190\,$\pm$\,3 [0] & 196\,$\pm$\,9$^\dagger$ [5] & 203\,$\pm$\,5$^\dagger$ [0] & 161\,$\pm$\,15$^\dagger$ [0] & 104\,$\pm$\,14 [17] & 112\,$\pm$\,0 [17] \\
\midrule
EDMD regulator, clipped & 187 [0] & 121 [37] & 188 [0] & 179 [5] & 117 [11] & 112 [16] \\
EDMD regulator, box QP & 187 [0] & 163 [21] & 188 [0] & 179 [5] & n/a & n/a \\
Benchmark PID / LQR & 3 [100] & 24 [100] & 10 [100] & 6 [100] & 115 [11] & 64 [21] \\
\bottomrule
\end{tabular}
}
\end{table}

\begin{figure}[t]
\centering
\includegraphics[width=0.80\textwidth]{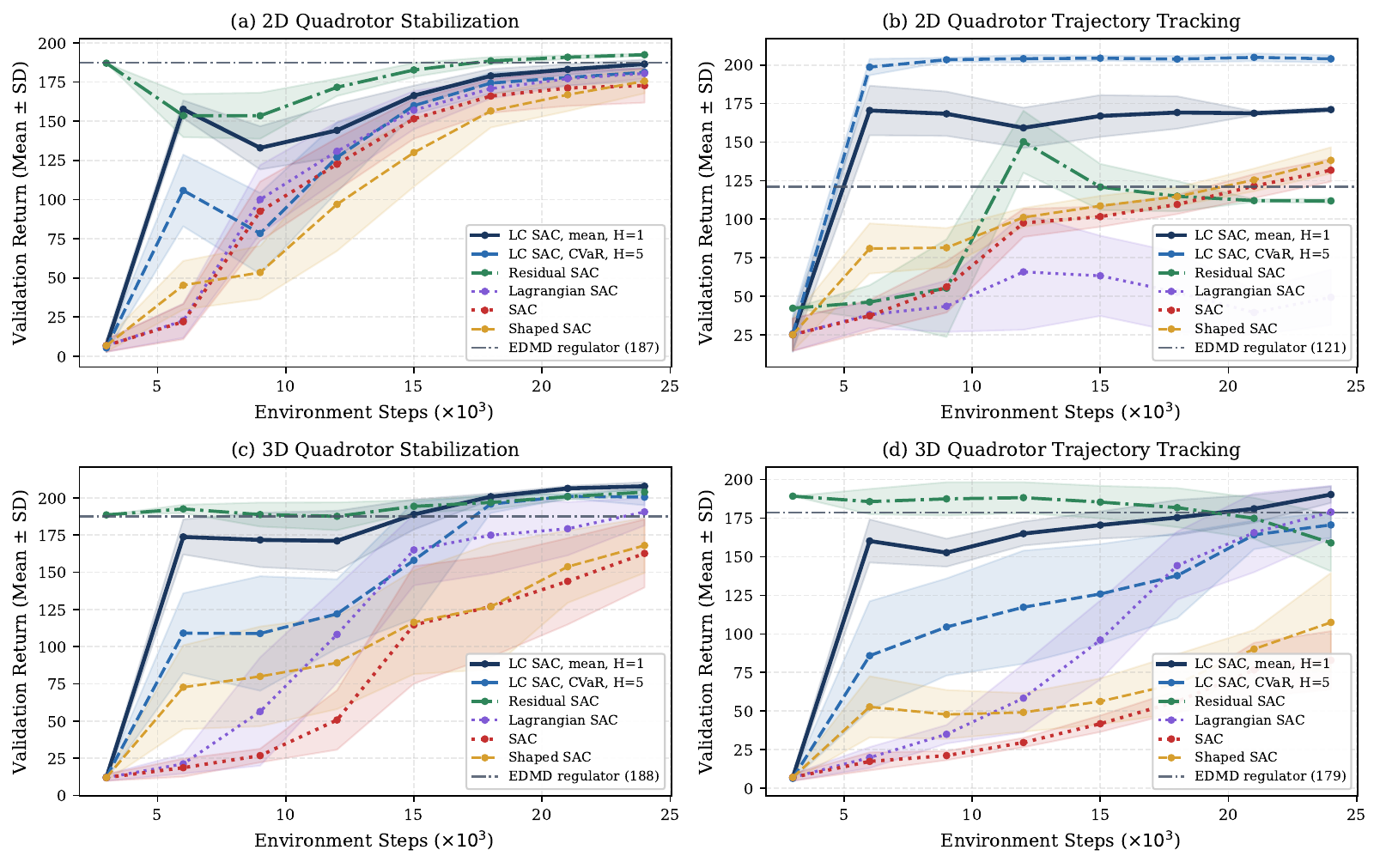}
\caption{Validation return during training (mean $\pm$ SD over seeds, with ten seeds except five for CVaR with $H{=}5$). The horizontal line is the box clipped EDMD regulator on the certification episodes. Residual SAC starts from the regulator scored on the six validation episodes, which differ. The headline variant learns tracking within about 6{,}000 steps on both quadrotors. Residual SAC peaks early on the tracking tasks and then declines.}
\label{fig:curves}
\end{figure}

\paragraph{At the nominal box:} Table~\ref{tab:nominal} and Figure~\ref{fig:curves} show the nominal results. The headline variant improves on SAC in mean on all four quadrotor tasks, significantly on three, and over the four tasks it beats SAC on a random pair of seeds with probability \piAllHead{} [\piAllHeadLo, \piAllHeadHi]. On 2D tracking it reaches \nomTwoTrackHead{} with \nomTwoTrackHeadFail\% failed test episodes, against \nomTwoTrackSac{} (\nomTwoTrackSacFail\%) for SAC and \nomTwoTrackRes{} (\nomTwoTrackResFail\%) for residual SAC on the EDMD regulator (Welch $p=\pTwoTrackHeadVsRes$). Residual SAC on the saturation aware regulator reaches \nomTwoTrackResQp{} ($p=\pTwoTrackHeadVsResQp$), and the multistep variants up to \nomTwoTrackBest{} with five seeds. Its failures come from one test start 1.3\,cm below the upper bound of the state space, failed on every seed, while those of SAC come from two seeds that leave the upper bound late in nearly every episode. On 3D tracking it reaches \nomThreeTrackHead{} against \nomThreeTrackSac{} for SAC (\nomThreeTrackSacFail\% failures) and \nomThreeTrackRes{} for residual SAC ($p=\pThreeTrackHeadVsRes$). Validation selection on six episodes favors residual SAC, which peaks early and then declines (Figure~\ref{fig:curves}). At the last checkpoint, the headline variant leads on both tracking tasks (\finTwoTrackHead{} against \finTwoTrackRes{} and \finThreeTrackHead{} against \finThreeTrackRes, both $p<0.001$). The Lagrangian SAC, which enforces the same hinge at the executed action through a cost critic, does well on 3D tracking (\nomThreeTrackLag) but collapses on 2D tracking (\nomTwoTrackLag, \nomTwoTrackLagFail\% failures). Domain randomized SAC (\nomTwoTrackDr{} and \nomThreeTrackDr) and shaped SAC (\nomTwoTrackShape{} and \nomThreeTrackShape) stay close to SAC, so the information in $W$ given as a reward does not close the gap. The action prior reaches \ablTwoTrackPrior{} and \ablThreeTrackPrior: the headline variant is ahead on 2D tracking by \ablTwoTrackPriorDiff{} \ablTwoTrackPriorCI{}, which is not significant (Holm $p=\ablTwoTrackPriorHolm$), and trails on 3D tracking, $\ablThreeTrackPriorDiff$ \ablThreeTrackPriorCI{} (Holm $p=\ablThreeTrackPriorHolm$). At the nominal box, then, the penalty is not a better use of the model than a tuned prior or a residual policy on 3D tracking. Learning curve areas and the distance to the state bounds are in Appendix~\ref{app:auc}.

\paragraph{Stabilization and the fixed regulators:} Residual SAC reaches \nomTwoStabRes{} and \nomThreeStabRes{} on 2D and 3D stabilization, against \nomTwoStabHead{} and \nomThreeStabHead{} for the headline variant ($p=\pThreeStabHeadVsRes$ on 3D), and the fixed regulator alone \nomTwoStabLqr{} and \nomThreeStabLqr{}. The headline variant beats SAC (\nomTwoStabSac{} and \nomThreeStabSac) significantly on 3D stabilization only, so on 2D stabilization the model is better used as the base of a residual policy. On the 19 certification episodes of 2D tracking the box clipped regulator reaches only \nomTwoTrackLqr{} (\nomTwoTrackLqrFail\% failures), the regulator that solves the box QP exactly \nomTwoTrackQp{} (\nomTwoTrackQpFail\%), and the headline variant \tcTwoTrackHead{} (\tcTwoTrackHeadFail\%). On 3D tracking, the two regulators coincide at \nomThreeTrackLqr, against \tcThreeTrackHead{} for the headline variant.

\paragraph{When the plain target is unattainable:} Replaying the states visited during training through the exact QP, the plain target $(1-\eta)W(h)$ lies below $W^\ast_k(h)$, so that no admissible action meets it, on \feasTwoTrackHeadPostMin{} to \feasTwoTrackHeadPostMax\% of the 2D tracking states visited by LC SAC in each phase after warm up and on \feasTwoTrackSacPostMin{} to \feasTwoTrackSacPostMax\% of those visited by SAC. With the plain threshold the aggregate stays above its tolerance on \gposTwoTrackAbsMin\% of the updates after warm up and the multiplier rises to about \multTwoTrackAbsMax{} and stays there, while with the relative threshold it is at most \multTwoTrackHeadMax{} as a phase average. The corresponding test returns are \nomTwoTrackAbs{} against \nomTwoTrackHead{} ($p=\pTwoTrackHeadVsAbs$). The plain threshold also calibrates its tolerance from its own hinge (median $\zeta$ of 0.0077 against 0.0029 on 2D tracking), so the ablation changes the scale of the constraint as well as its target. On the other tasks the plain target becomes attainable after warm up on all but at most \feasOtherHeadMax\% of the visited states. The plain threshold gives lower returns on stabilization ($p=\pTwoStabHeadVsAbs$ and $\pThreeStabHeadVsAbs$), where it is too tight rather than infeasible, and not on 3D tracking ($p=\pThreeTrackHeadVsAbs$). The share alone does not predict the gain (below).

\paragraph{Perturbations:} Table~\ref{tab:robust} in the appendices evaluates the same checkpoints under unseen changes of mass, inertia and wind (R1 to R4). The perturbations that push the vehicle down remove most 2D tracking failures, while with 5\% less mass every learner often leaves the upper bound, the headline variant in \robRfourTwoTrackHeadFail\% of episodes against \robRfourTwoTrackResFail\% for residual SAC. No learner avoids failures in both directions, and the constraint does not help with this.

\begin{figure}[t]
\centering
\includegraphics[width=\textwidth]{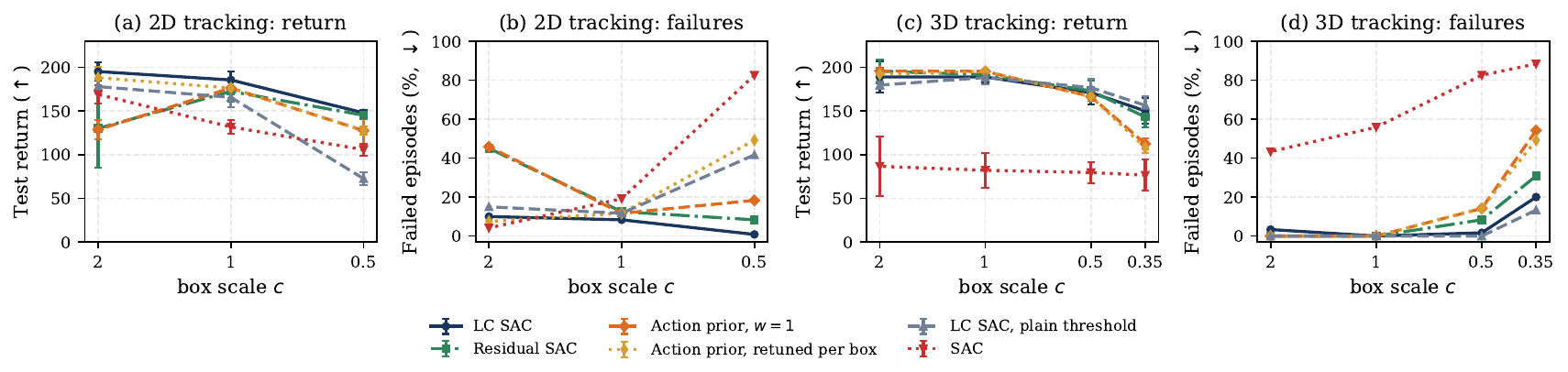}
\caption{Varying the input box: the policy action $a\in[-1,1]^m$ commands $v=c\,a$, and every learner is retrained at each $c$ with the same model and $W$. Test return (mean $\pm$ SD over seeds) and failed test episodes are shown for ten seeds, except five on 3D tracking at $c=2$, and for the action prior at $c=1$ on 3D tracking. The action prior uses $w=1$, selected at the nominal box, or the weight selected again at each box. The box shrinks to the right.}
\label{fig:box}
\end{figure}

\paragraph{Across input boxes:} To ask whether the way the model is given to SAC survives a change of the input limits, we scale the input box by $c$, retrain every learner at each $c$, and keep the model, $W$ and every setting fixed, including the weight of the action prior (Figure~\ref{fig:box} and Table~\ref{tab:box}). The learners also differ in how they meet a new box. LC SAC sets its tolerance $\zeta$ from the hinge of the feedforward action on the warm up buffer of each run (Section~\ref{sec:constraint}), so the scale of its constraint follows the box without retuning, while the scale of the residual and the entropy temperature of residual SAC are not tuned at any box. The model is linear in $v$, so $c=2$ extrapolates it beyond the inputs it was identified on. The thrust limits of the vehicles do not bind at any $c$, so the sweep changes the control authority around the feedforward, not the hardware limits. The headline variant has higher mean return than SAC at all \boxNpoints{} evaluated pairs of task and box, by \boxGainSacMin{} to \boxGainSacMax{} in return (Holm $p\le\boxPSacMax$). In the smallest boxes, SAC fails in \boxTwoTrackCHalfSacFail\% and \boxThreeTrackCLowSacFail\% of test episodes against \boxTwoTrackCHalfHeadFail\% and \boxThreeTrackCLowHeadFail\%, although in the largest 2D box it fails less (\boxTwoTrackCTwoSacFail\% against \boxTwoTrackCTwoHeadFail\%). At every pair its mean return is within \worstGapHead{} points of the best mean among SAC, residual SAC, the action prior with fixed or retuned weight, and the plain threshold. The learners anchored on the regulator are not: residual SAC breaks down in the largest 2D box (\boxTwoTrackCTwoRes, \boxTwoTrackCTwoResFail\% failures, although its base regulator alone reaches \boxTwoTrackCTwoQp{} there), and so does the action prior (\boxTwoTrackCTwoPrior, \boxTwoTrackCTwoPriorFail\%), which also falls behind in the smallest boxes (\boxTwoTrackCHalfPrior{} and \boxThreeTrackCLowPrior, with \boxTwoTrackCHalfPriorFail\% and \boxThreeTrackCLowPriorFail\% failures). The plain threshold falls to \boxTwoTrackCHalfAbs{} in the smallest 2D box. On 3D tracking, it is never significantly different from the headline variant. On 3D tracking residual SAC is never significantly different from the headline variant, and in the largest box both anchored learners are a few points higher in mean. Shrinking the box raises the share of states on which the plain target is unattainable, from \boxTwoTrackCTwoHeadShare\% to \boxTwoTrackCHalfHeadShare\% on 2D tracking and from \boxThreeTrackCTwoHeadShare\% to \boxThreeTrackCLowHeadShare\% on 3D tracking, but the gain over SAC does not grow with it. On both tasks the gain is largest at $c=1$.

\begin{table}[H]
\centering
\caption{Input box sweep (Figure~\ref{fig:box}): test return (mean $\pm$ SD over seeds) with failed test episodes in \% in brackets. Results use ten seeds, except five on 3D tracking at $c=2$, and for the action prior at $c=1$ on 3D tracking. The action prior uses $w=1$, selected on three 2D tracking training seeds at the nominal box. The retuned prior uses the weight in parentheses, selected on validation return at each box other than the nominal one. At $c=1$ we use the nominal selection $w=1$, made on 2D tracking. $^\dagger$: LC SAC better than SAC, $^\ddagger$: better than residual SAC, both Welch with Holm correction over $c$ within the task.}
\label{tab:box}
\resizebox{\textwidth}{!}{
\begin{tabular}{llrrrrrr}
\toprule
Task & $c$ & LC SAC & SAC & Residual & Prior & Retuned prior ($w$) & Plain target \\
\midrule
2D track. & 2 & 195\,$\pm$\,10$^\dagger$$^\ddagger$ [10] & 169\,$\pm$\,10 [4] & 130\,$\pm$\,45 [45] & 129\,$\pm$\,11 [46] & 188\,$\pm$\,12 [8] (0.1) & 178\,$\pm$\,8 [15] \\
2D track. & 1 & 186\,$\pm$\,10$^\dagger$$^\ddagger$ [8] & 132\,$\pm$\,8 [19] & 172\,$\pm$\,12 [12] & 176\,$\pm$\,11 [12] & 176\,$\pm$\,11 [12] (1) & 166\,$\pm$\,12 [12] \\
2D track. & 0.5 & 148\,$\pm$\,4$^\dagger$ [1] & 106\,$\pm$\,6 [82] & 145\,$\pm$\,4 [8] & 127\,$\pm$\,16 [18] & 128\,$\pm$\,5 [49] (0.1) & 72\,$\pm$\,7 [42] \\
3D track. & 2 & 189\,$\pm$\,18$^\dagger$ [3] & 87\,$\pm$\,34 [43] & 196\,$\pm$\,14 [0] & 196\,$\pm$\,4 [0] & 192\,$\pm$\,3 [0] (0.1) & 180\,$\pm$\,5 [0] \\
3D track. & 1 & 189\,$\pm$\,6$^\dagger$ [0] & 82\,$\pm$\,20 [56] & 192\,$\pm$\,2 [0] & 195\,$\pm$\,1 [0] & 195\,$\pm$\,1 [0] (1) & 188\,$\pm$\,8 [0] \\
3D track. & 0.5 & 171\,$\pm$\,14$^\dagger$ [2] & 79\,$\pm$\,12 [82] & 173\,$\pm$\,11 [8] & 167\,$\pm$\,4 [14] & 167\,$\pm$\,4 [14] (1) & 177\,$\pm$\,10 [0] \\
3D track. & 0.35 & 150\,$\pm$\,14$^\dagger$ [20] & 76\,$\pm$\,18 [88] & 143\,$\pm$\,12 [31] & 112\,$\pm$\,7 [54] & 108\,$\pm$\,6 [49] (10) & 156\,$\pm$\,10 [13] \\
\bottomrule
\end{tabular}
}
\end{table}

\paragraph{Is the prior only mistuned?} A fixed $w$ is a different weight in input units at each $c$ (Section~\ref{sec:constraint}), so we also selected $w$ at every box, by the same rule on validation return over three seeds. Retuning rescues the prior in the largest 2D box (\boxTwoTrackCTwoPriorT{} with $w=\boxTwoTrackCTwoPriorW$, a gap to the headline variant of \retunedTwoTrackCTwoGain{} \retunedTwoTrackCTwoCI, Holm $p=\retunedTwoTrackCTwoP$), but not in the two smallest boxes, where the retuned prior still trails by \retunedTwoTrackCHalfGain{} \retunedTwoTrackCHalfCI{} on 2D tracking and \retunedThreeTrackCLowGain{} \retunedThreeTrackCLowCI{} on 3D tracking (Holm $p=\retunedTwoTrackCHalfP$ and $\retunedThreeTrackCLowP$ over the three boxes tested) and fails in \boxTwoTrackCHalfPriorTFail\% and \boxThreeTrackCLowPriorTFail\% of episodes. Where the selected weight was at an edge of this grid, $w=0.1$ at three boxes and $w=10$ in the smallest 3D box, we also ran the next weight outward, $w=0.01$ or $w=100$, on the same three seeds. The additional weight scored lower in validation at all four boxes, so every selected weight is an interior best of the weights tried. At each box other than the nominal one the selection used nine runs, three weights on three seeds, of which the three runs with $w=1$ came from the sweep, plus the three runs of the extension at the four edge boxes. When the selected weight was not one, we ran it on the remaining seeds. The headline variant used one setting throughout.

\paragraph{The radius of the zero penalty set:} Table~\ref{tab:ablate} compares variants at the nominal box, and Table~\ref{tab:box_diagnostics} gives the floor radius sweep. Holding the radius at its slack floor, $\tau_k=W^\ast_k+\epsilon_\tau(W+w_0)$, reduces return on 2D tracking by \boxTwoTrackCTwoGainFloor, \boxTwoTrackCOneGainFloor{} and \boxTwoTrackCHalfGainFloor{} points at $c=2$, 1 and 0.5 (Holm $p=\boxTwoTrackCTwoPFloor$, $\boxTwoTrackCOnePFloor$ and $\boxTwoTrackCHalfPFloor$). In the largest box every seed of the floor variant fails the same nine of the twelve test starts within a few steps of each other, hence a standard deviation near zero. With the radius at its floor, the penalty leaves the actor little room around the greedy model action $v^\ast_k$, which has no input cost and is aggressive when the box is large. The loss is largest in the largest box and not significant in the smallest, where the plain target is unattainable on \boxTwoTrackCHalfHeadShare\% of the states the headline variant visits, so its radius is already at the floor there. On 3D tracking none of these differences is significant (\boxThreeTrackCTwoGainFloor, \boxThreeTrackCOneGainFloor{} and \boxThreeTrackCLowGainFloor{} at $c=2$, 1 and 0.35). Negative values favor the floor variant. A zero penalty set wider than the slack floor helps on 2D tracking, but we see no such effect on 3D tracking.

\paragraph{An isotropic metric:} The isotropic surrogate replaces the predicted excess by $\mu\|v-v^\ast_k\|^2$, with $\mu$ the mean eigenvalue of $G$. It reduces nominal return by \ablTwoTrackEuclidDiff{} and \ablThreeTrackEuclidDiff{} points ($p=\ablTwoTrackEuclidP$ and $\ablThreeTrackEuclidP$). By Eq.~\eqref{eq:gap} the replacement also drops the linear boundary term, and its tolerance is recalibrated from the new hinge, so this ablation does not separate the metric from the boundary term.

\paragraph{Simpler thresholds and multipliers:} At the nominal box, we detect no return difference for the backup threshold (\ablTwoTrackBackup{} and \ablThreeTrackBackup) or a fixed multiplier with the quadratic term (\ablTwoTrackFixlam{} and \ablThreeTrackFixlam), against \ablTwoTrackHead{} and \ablThreeTrackHead{} for the headline variant. We ran these two variants only at the nominal box, and the fixed value was read off the adaptive runs. Without the quadratic term it reaches \ablTwoTrackFixlamRho{} on 2D tracking, but falls to \ablThreeTrackFixlamRho{} on 3D tracking, where the chosen multiplier is small.

\begin{table}[H]
\centering
\caption{Variants of the penalty at the nominal box. The isotropic surrogate also drops the boundary term of Eq.~\eqref{eq:gap}. Test return (mean $\pm$ SD over seeds) with failed test episodes in \% in brackets. We use ten seeds on 2D tracking and ten or five on 3D tracking as marked.}
\label{tab:ablate}
{\footnotesize
\begin{tabular}{lll}
\toprule
Variant at $c=1$ & 2D tracking & 3D tracking \\
\midrule
LC SAC (relative threshold, adaptive dual) & 186\,$\pm$\,10 [8] & 189\,$\pm$\,6 [0] \\
backup threshold (clipped regulator) & 184\,$\pm$\,7 [8] & 184\,$\pm$\,5 [0] (5 seeds) \\
fixed multiplier & 190\,$\pm$\,4 [8] & 192\,$\pm$\,4 [0] (5 seeds) \\
fixed multiplier, $\rho_a=0$ & 187\,$\pm$\,3 [8] & 121\,$\pm$\,11 [3] (5 seeds) \\
Isotropic surrogate around $v^\ast$ & 170\,$\pm$\,8 [2] & 166\,$\pm$\,9 [0] (5 seeds) \\
radius at the slack floor & 157\,$\pm$\,1 [25] & 187\,$\pm$\,2 [0] (5 seeds) \\
plain threshold & 166\,$\pm$\,12 [12] & 188\,$\pm$\,8 [0] \\
action prior toward the regulator ($w=1$) & 176\,$\pm$\,11 [12] & 195\,$\pm$\,1 [0] (5 seeds) \\
\bottomrule
\end{tabular}
}
\end{table}

\paragraph{What the certificate says:} Table~\ref{tab:closedloop} in the appendices gives the median $\hat\rho$ over seeds. On 2D stabilization every learner except LAC, including SAC, is certified with $\hat\rho\approx 0.98$ on every seed. On the tracking tasks $\hat\rho$ is at least one on every seed of every method except three seeds on 3D tracking, at 0.998 and 0.999, and on 2D tracking most LC SAC seeds fail certification starts 3 to 11\,cm below the upper bound, so their $\hat\rho$ is infinite while the median for SAC is 2.8. The score takes the worst step of the worst episode, and the constraint acts on model predictions over replay states, so we read $\hat\rho$ as a diagnostic available for every controller, not as evidence of stability.

\paragraph{Other variants:} The horizon matters more than the aggregate. $H=5$ is the best choice on 2D tracking (206 against 184 for $H=1$ with CVaR), while $H=1$ is best on 3D tracking (186 against 169 and 162 for $H=5$ and $H=10$), where longer predictions of the linear model are less accurate. Mean and CVaR aggregation are within one standard deviation on most tasks, with no advantage of CVaR in return, and the adaptive quantile behaves like a fixed $q\approx 0.9$. LAC, with its own cost and architecture, does not learn these tasks within 24{,}000 steps. On the cartpole, SAC, every LC SAC variant we ran and the EDMD regulator score between 104 and 117 of 150 with 8 to 17\% failures, and the median certificate of every learner is infinite, so these tasks do not separate the methods. Changing $c_\zeta$ or $\rho_a$ by a factor of three in either direction, on 2D tracking at the nominal box with three seeds, changes the mean test return of the headline variant by at most 10 points, less than one standard deviation over the ten seeds of the main runs, and does not change the failure rate. The looser tolerance $c_\zeta=0.3$ lowers the return by 9.6 on each of the three seeds. Each of the other three changes shifts the return by at most 2.2 points.

\section{Limitations}
\label{sec:limitations}

\paragraph{Training budget:} We study early learning over 24{,}000 steps. Figure~\ref{fig:curves} suggests that SAC is still improving at that point. Whether it catches up with a larger budget remains open. Besides these steps, every method that uses the model relies on 24 identification episodes, and model selection, validation and certification use further episodes.

\paragraph{No guarantee on the plant:} The penalty is elastic and acts on replay states, so the trained policy is not guaranteed to satisfy it, and nothing here is a stability guarantee on the plant.

\paragraph{Scope of the box sweep:} The sweep covers two tracking tasks with one model each, scaling the box above one extrapolates that model, and the entropy temperature is fixed, so exploration noise in input units grows with $c$ for every learner. The floor radius effect is significant on 2D tracking only, the isotropic surrogate changes both metric anisotropy and the boundary term, and at the nominal box a tuned action prior is better on 3D tracking.

\paragraph{Tuning:} The weight of the action prior was selected on validation return, from four values at the nominal box and three or four at the others, and the fixed multiplier was read off the adaptive runs. No other setting was tuned per task, and the sensitivity of LC SAC was tested only for $c_\zeta$ and $\rho_a$, on 2D tracking. The Lagrangian SAC applies the tolerance matched for the deterministic hinge to the noisier hinge of the executed action.

\paragraph{Perturbations and simulation:} The perturbations are within a small range of mass, inertia and wind, and every learner is sensitive to their direction, in part because the 2D circle touches the upper bound of the state space. Everything is in simulation.

\paragraph{Cost:} The headline variant takes a median of 1.3 and 2.4 times the wall time of SAC per run on 2D and 3D tracking, and the variant with $H=5$ up to 3.7 times (Appendix~\ref{app:compute}).

\section{Conclusion}
\label{sec:conclusion}
A Lyapunov decrease target that an admissible action can always meet turns a data driven Riccati certificate into an actor penalty for SAC. The penalty is zero on a sublevel set of the predicted quadratic around the best admissible model action, whose radius cannot shrink when the input box grows. With one setting, the penalty had a higher mean return than SAC at every pair of tracking task and input box we tried and stayed within \worstGapHead{} return points of the best compared method at each. On 2D tracking, residual SAC, an action prior toward the same regulator and a plain contraction target each fell far behind at some box, while on 3D tracking residual SAC was not significantly different from the penalty. Retuning the prior at every box did not close the gap in the smallest boxes. Holding the zero penalty set at its slack floor costs return on 2D tracking, while the backup threshold and a fixed multiplier make no detectable difference at the nominal box. At the nominal box a tuned action prior is better on 3D tracking, no learner is robust to unseen perturbations in both directions, and the closed loop certificate shows that the policies do not contract $W$ at every step on the tracking tasks. Every learner was retrained at each box, and everything is in simulation. Certifying the decrease on the plant is the next step.

\paragraph{Reproducibility:} The code release will contain the identification data and collection code, the identified models, the controllers and learners, the configuration files, and the evaluation and aggregation scripts, with the commands for each stage. All results use the benchmark reward and the deployed policies without any filter. Seed level intervals are conditional on the fixed evaluation starts and on one identification dataset per task.

\clearpage
\bibliography{main}
\bibliographystyle{tmlr}

\clearpage
\appendix

\section{Proofs}
\label{app:proofs}

\paragraph{Proposition~\ref{prop:feasible}:} (a) $v\mapsto W(Ah+Bv)$ is a strictly convex quadratic with Hessian $2B^\top PB\succ 0$, so it has a unique minimizer on the compact convex box $\Ucal_k$. Since $\tau_k(h)\ge W^\ast_k(h)+\epsilon_\tau(W(h)+w_0)\ge W^\ast_k(h)$, the hinge vanishes at $v^\ast_k(h)$. Continuity of $h\mapsto v^\ast_k(h)$ follows from strict convexity and continuity of the problem data. (b) Let $\bar\Delta=\sup_{h,k}\big(W(Ah)-W^\ast_k(h)\big)$ over a compact set, and $\bar s=\min\{1,\epsilon_\tau w_0/\bar\Delta\}$ ($\bar s=1$ if $\bar\Delta=0$). By convexity, $W(\hat h^+((1-\bar s)v^\ast))\le(1-\bar s)W^\ast_k+\bar s\,W(Ah)\le W^\ast_k+\bar s\bar\Delta\le W^\ast_k+\epsilon_\tau w_0\le\tau_k$, and $(1-\bar s)v^\ast$ lies in the interior because the origin does. (c) With the action map $v=c\,a$, the map $o\mapsto\mathrm{arctanh}((1-\bar s)v^\ast_{k}(h(o))/c)$ is continuous on a compact set and bounded, since the selection stays in the interior, so a ReLU network approximates it uniformly \citep{leshno1993multilayer}. The hinge is Lipschitz in $a$ on this set, which gives $\sup_o\ell<\zeta$ for a close enough approximation. Both the mean and the CVaR of a variable bounded by $\zeta'<\zeta$ are at most $\zeta'$. \hfill$\square$

\paragraph{Proposition~\ref{prop:deadzone}:} Since $q$ is quadratic with Hessian $2G$, $q(v)=q(v^\ast_k)+\nabla q(v^\ast_k)^\top(v-v^\ast_k)+\|v-v^\ast_k\|_G^2$, and $q(v^\ast_k)=W^\ast_k$. As $v^\ast_k$ minimizes the convex $q$ over the convex box, $\nabla q(v^\ast_k)^\top(v-v^\ast_k)\ge 0$ for every $v$ in the box, with $\nabla q(v^\ast_k)=0$ if $v^\ast_k$ is interior. The hinge is zero iff $q(v)\le\tau_k=W^\ast_k+r_k$, which by the expansion implies $\|v-v^\ast_k\|_G^2\le r_k$, with equivalence on $\Ucal_k$ for interior $v^\ast_k$. The unrestricted ellipsoid must still be intersected with the box. A larger box contains a smaller one, so the minimum $W^\ast_k$ cannot increase with $c$, and $r_k=\max\{t-W^\ast_k,\delta\}$ cannot decrease, since $t$ does not depend on $c$. If $(1-\eta)W\ge W^\ast_k+\epsilon_\tau(W+w_0)$, then $\tau_k=(1-\eta)W$ does not depend on $c$. The slack floor is active when $(1-\eta)W-W^\ast_k\le\epsilon_\tau(W+w_0)$, regardless of whether a bound is active at the minimizer. Finally $\|v-v^\ast_k\|_G^2=c^2\|a-v^\ast_k/c\|_G^2$ for $v=c\,a$. \hfill$\square$

\paragraph{Proposition~\ref{prop:cvar}:} The dual representation is $\CVaR_q^\mu(\ell)=\sup\{\mathbb{E}_\mu[\ell\xi]: 0\le\xi\le 1/(1-q),\ \mathbb{E}_\mu\xi=1\}$ \citep{rockafellar2002conditional}. Taking $\xi\equiv 1$ gives $\mathbb{E}_\mu\ell\le\CVaR_q$. Bounding $\xi\le 1/(1-q)$ and using $\ell\ge 0$ gives $\CVaR_q\le\mathbb{E}_\mu\ell/(1-q)$, which proves (a). For (b), $w=d\nu/d\mu\le 1/(1-q)$ and $\mathbb{E}_\mu w=1$, so $\xi=w$ is feasible in the dual and $\mathbb{E}_\nu\ell=\mathbb{E}_\mu[\ell w]\le\CVaR^\mu_q(\ell)$. \hfill$\square$

\paragraph{Proposition~\ref{prop:conformal}:} Each episode is a deterministic function of its initial state, so the $n+1$ scores are i.i.d. and hence exchangeable. By exchangeability the new score exceeds the $\lceil(n+1)(1-\alpha)\rceil$ smallest of the $n+1$ scores with probability at most $\alpha$, and ties, as at $+\infty$, only make this conservative, so $\mathbb{P}(s_{n+1}\le\hat\rho)\ge 1-\alpha$ \citep{vovk2005algorithmic}. On this event and if $\hat\rho<\infty$, the score is finite, so the episode did not fail, and $W(h_{t+1})-\omega\le s_{n+1}W(h_t)\le\hat\rho\,W(h_t)$ for all $t$. Iterating gives the bound. \hfill$\square$

\section{Dual variable}
\label{app:dual}

\begin{proposition}[Bounded and relaxing dual]
Let $\lambda_{n+1}=[(1-\beta\epsilon)\lambda_n+\beta g_n]_+$ with $0<\beta\epsilon<1$ and $\lambda_0\ge 0$, and let $\bar g=\sup_n[g_n]_+$. (a) $\lambda_n\le\max\{\lambda_0,\bar g/\epsilon\}$ for all $n$. (b) If $g_n\le-\kappa<0$ for all $n\ge n_0$, then $\lambda_n=0$ for all $n\ge n_0+\lceil\lambda_{n_0}/(\beta\kappa)\rceil$. (c) If $g_n\equiv g$, the fixed point is $\lambda^\ast=[g]_+/\epsilon$.
\end{proposition}

\begin{proof}
(a) Let $M=\max\{\lambda_0,\bar g/\epsilon\}$. If $\lambda_n\le M$, then $\lambda_{n+1}\le(1-\beta\epsilon)M+\beta\bar g\le(1-\beta\epsilon)M+\beta\epsilon M=M$, and induction gives the claim. (b) For $n\ge n_0$, $\lambda_{n+1}\le[\lambda_n-\beta\kappa]_+$, so $\lambda$ decreases by at least $\beta\kappa$ per update until it reaches zero, where it stays. (c) Solve $\lambda=(1-\beta\epsilon)\lambda+\beta g$ for $g>0$. For $g\le 0$, the map has the fixed point zero.
\end{proof}

At the fixed point $\lambda^\ast=[g]_+/\epsilon$ and $\lambda$ is held constant during the actor step, so the actor gradient is $\nabla(\text{SAC})+(\lambda^\ast+\rho_a[g]_+)\nabla g$, the gradient of the SAC objective plus $\big(\frac{1}{2\epsilon}+\frac{\rho_a}{2}\big)[g]_+^2$, a quadratic penalty on the relative violation. This calculation characterizes stationary points. Convergence of the iterates remains open. The constraint is therefore elastic: it allows a residual violation that decreases with $1/\epsilon$, and it does not impose $C\le\zeta$ exactly.

\section{Implementation Details}
\label{app:impl}

\subsection{Algorithm}
\label{app:alg}

\begin{algorithm}[!htbp]
\caption{LC SAC update (one gradient step)}
\label{alg:lcsac}
\begin{algorithmic}[1]
\State Sample a batch $\{(o_i,a_i,r_i,o_i',e_i,k_i,\bar u_{k_i})\}$ from replay; $h_i\gets\psi(S^{-1}e_i)$.
\State Update the twin critics and their targets as in SAC.
\State $a^{\mathrm{det}}_i\gets\tanh\mu_\theta(o_i)$; \ $v_i\gets D_u^{-1}(\mathrm{clip}(\bar u_{k_i}+c\,D_u a^{\mathrm{det}}_i)-\bar u_{k_i})$.
\State Without gradient: $W^\ast_i$ by the exact box QP (or the $H$ step variant), $\tau_i$ by \eqref{eq:threshold}.
\State $\ell_i$ by \eqref{eq:hinge}; \ $C\gets\mathrm{mean}_i\,\ell_i$ or $\CVaR_q(\ell_i)$; \ $g\gets(C-\zeta)/\zeta$.
\State Actor step on $L_{\mathrm{SAC}}(\theta)+\lambda g+\frac{\rho_a}{2}[g]_+^2$, keeping the temperature fixed at 0.2.
\State $\lambda\gets[(1-\beta\epsilon)\lambda+\beta\,g]_+$ with $g$ detached.
\end{algorithmic}
\end{algorithm}

\subsection{Settings}
\label{app:settings}

\begin{table}[H]
\centering
\caption{Settings shared by all tasks. The LC SAC settings were fixed before any run reported here and never changed ($q$ was carried over from an earlier version of the method): $\eta=\gamma/2$ lies inside the contraction of the regulator, $\epsilon_\tau$ and $w_0$ only keep the threshold strictly feasible and the hinge finite near zero error ($w_0$ is small against the mean of $W$, which is one on the identification data), and $c_\zeta$, $q$, $\beta$, $\epsilon$ and $\rho_a$ were set without a search. The only pilot (three seeds, 16{,}000 steps) compared the horizons $H\in\{1,5,10\}$. Selected on these tasks: only the multiplier of the fixed multiplier ablation, read off the adaptive runs, and the weight of the action prior, selected on validation return.}
\label{tab:settings}
\small
\begin{tabular}{@{}lp{0.68\textwidth}@{}}
\toprule
SAC (all SAC based learners) & twin critics, two hidden layers of 128 ReLU units, Adam with learning rate $3\cdot 10^{-4}$ for actor and critics, batch 256, discount 0.99, Polyak rate 0.005, entropy temperature fixed at 0.2, Huber critic loss, gradient norms clipped at 10, replay of $10^6$ transitions, one update per environment step, 3{,}000 warm up steps with uniform random actions\\
LC SAC & $\eta=\gamma/2$, $\epsilon_\tau=10^{-3}$, $w_0=10^{-2}$, $c_\zeta=0.1$, $q=0.75$, dual step $\beta=10^{-3}$, regularization $\epsilon=0.1$, $\rho_a=1$\\
Adaptive quantile & $q_t\in[0.5,0.9]$, $\sigma=0.25$\\
Shaped SAC & weight 0.5 on $W(h_k)-0.99\,W(h_{k+1})$\\
Lagrangian SAC & cost critic on the hinge of the executed action, PI gains 0.1 and $10^{-3}$, tolerance $\zeta$ as for the mean aggregate\\
Action prior & $w\|a-v_b/c\|^2$ at the deterministic action, $v_b=\mathrm{clip}(-Kh,\Ucal_k)$; $w\in\{0.1,1,10,100\}$ by validation at the nominal box ($w=1$), $\{0.1,1,10\}$ at the other boxes, extended to $0.01$ or $100$ where the selection was at an edge\\
Residual SAC & base action $\mathrm{clip}(-Kh,\Ucal_k)$, or the box QP regulator (2D tracking), actor output layer initialized at zero\\
Domain randomized SAC & mass scale $[0.97,1.03]$, inertia scale $[0.90,1.10]$, wind scale $[0,0.5]$ redrawn at each reset\\
LAC & hidden layers of 64 units, discount 0.995, learning rates $10^{-4}$ (actor and temperature) and $3\cdot 10^{-4}$ (critic), $\alpha=2$\\
\bottomrule
\end{tabular}
\end{table}

\subsection{Selected Models}
\label{app:models}

\begin{table}[H]
\centering
\caption{Selected models. $\gamma$: per step contraction of $W$ under the unsaturated regulator. Last columns: worst leave one episode out fold of $\rho_B/\varepsilon_P$ for the earlier and the final collection.}
\label{tab:certificate}
{\footnotesize
\begin{tabular}{llrrrrrr}
\toprule
Task & model & $n_h$ & $\gamma$ & $\rho(A-BK)$ & $\mathrm{cond}(P)$ & \multicolumn{2}{c}{worst fold $\rho_B/\varepsilon_P$} \\
 & & & & & & earlier & final \\
\midrule
2D stab. & lin & 6 & 0.0136 & 0.987 & 106 & 3.07 & 35.97 \\
2D track. & lin & 6 & 0.0141 & 0.986 & 221 & 0.95 & 3.60 \\
3D stab. & lin & 12 & 0.0125 & 0.989 & 121 & 0.35 & 24.23 \\
3D track. & lin & 12 & 0.0133 & 0.984 & 168 & 0.70 & 7.18 \\
Cartpole stab. & bil & 4 & 0.0353 & 0.926 & 48 & 2.78 & 2.78 \\
Cartpole track. & lin & 4 & 0.0360 & 0.929 & 46 & 1.14 & 1.14 \\
\bottomrule
\end{tabular}
}
\end{table}

\subsection{Compute}
\label{app:compute}

Training ran on one workstation with a 16 core AMD Ryzen 9 9950X CPU and one NVIDIA RTX 4000 Ada GPU, with up to ten runs in parallel sharing the GPU. Both the simulator and LAC training run on the CPU. A run of 24{,}000 steps took a median of 21 and 14 minutes for SAC on 2D and 3D tracking, 27 and 33 minutes for the headline variant, and 42 and 51 minutes for the variant with $H=5$, over all boxes and under the load of the parallel runs. The quadrotor learning runs reported here add up to about 350 hours, summed over runs.

\section{Additional Results}
\label{app:results}

\subsection{Further Results}
\label{app:extra}

\begin{table}[H]
\centering
\caption{Median over seeds of the closed loop certificate $\hat\rho$ ($n=19$, $\alpha=0.05$, $\omega=0.05$), with the number of seeds for which $\hat\rho<1$ in parentheses, for the baselines and three LC SAC variants. LAC is $\infty$ on every task. $\infty$: at least half of the seeds had a failed episode. On the cartpole every median is $\infty$.}
\label{tab:closedloop}
{\scriptsize
\begin{tabular}{lrrrr}
\toprule
Method & 2D stab. & 2D track. & 3D stab. & 3D track. \\
\midrule
SAC & 0.983 (10/10) & 2.815 (0/10) & 1.002 (4/10) & $\infty$ (0/10) \\
Shaped SAC & 0.980 (10/10) & 3.046 (0/10) & 1.017 (1/10) & $\infty$ (0/10) \\
Lagrangian SAC & 0.981 (10/10) & $\infty$ (0/10) & 0.990 (9/10) & 6.756 (0/10) \\
Domain rand. SAC & 0.982 (10/10) & 2.268 (0/10) & 1.000 (5/10) & $\infty$ (0/10) \\
Residual SAC & 0.981 (10/10) & $\infty$ (0/10) & 1.000 (5/10) & $\infty$ (1/10) \\
Residual SAC, QP base & n/a & $\infty$ (0/10) & n/a & n/a \\
\midrule
LC SAC, mean, $H{=}1$ & 0.981 (10/10) & $\infty$ (0/10) & 1.010 (2/10) & 4.872 (0/10) \\
LC SAC, plain threshold & 0.980 (10/10) & $\infty$ (0/10) & 0.988 (10/10) & 5.423 (2/10) \\
LC SAC, CVaR, $H{=}5$ & 0.980 (5/5) & $\infty$ (0/5) & 1.002 (2/5) & 9.261 (0/5) \\
\midrule
EDMD regulator, clipped & 0.976 & $\infty$ & 0.978 & $\infty$ \\
\bottomrule
\end{tabular}
}
\end{table}

\begin{table}[htbp]
\centering
\caption{Box sweep diagnostics. unatt.: share of the training states of the headline variant, every fourth state over all 24{,}000 steps with the warm up included, on which the plain target is unattainable. The text reports shares per phase after warm up. Floor radius: $\tau_k=W^\ast_k+\epsilon_\tau(W+w_0)$, ten seeds on 2D and five on 3D tracking. Box QP: the fixed saturation aware regulator, scored on the 19 certification episodes instead of the twelve test episodes. Markers as in Table~\ref{tab:box}.}
\label{tab:box_diagnostics}
\begin{tabular}{llrlll}
\toprule
Task & $c$ & unatt. & LC SAC & Floor radius & Box QP \\
\midrule
2D track. & 2 & 12\% & 195\,$\pm$\,10$^\dagger$$^\ddagger$ [10] & 70\,$\pm$\,0 [75] & 165 [26] \\
2D track. & 1 & 63\% & 186\,$\pm$\,10$^\dagger$$^\ddagger$ [8] & 157\,$\pm$\,1 [25] & 163 [21] \\
2D track. & 0.5 & 80\% & 148\,$\pm$\,4$^\dagger$ [1] & 143\,$\pm$\,6 [2] & 56 [63] \\
3D track. & 2 & 8\% & 189\,$\pm$\,18$^\dagger$ [3] & 175\,$\pm$\,7 [7] & 179 [5] \\
3D track. & 1 & 14\% & 189\,$\pm$\,6$^\dagger$ [0] & 187\,$\pm$\,2 [0] & 179 [5] \\
3D track. & 0.5 & 34\% & 171\,$\pm$\,14$^\dagger$ [2] & n/a & 168 [11] \\
3D track. & 0.35 & 50\% & 150\,$\pm$\,14$^\dagger$ [20] & 162\,$\pm$\,3 [8] & 136 [37] \\
\bottomrule
\end{tabular}

\end{table}

\subsection{Learning Speed and Distance to the Bounds}
\label{app:auc}

Table~\ref{tab:auc} gives the area under the validation curve, the mean validation return over the eight checkpoints from 3{,}000 to 24{,}000 steps, which weighs early learning as much as the final return. LC SAC has a larger mean validation return over these checkpoints than SAC on every quadrotor task. Comparisons with model informed baselines depend on the task. Residual SAC, which starts from the regulator, has the larger area on both stabilization tasks and on 3D tracking, but not on 2D tracking, where it peaks early and then declines (Figure~\ref{fig:curves}). The action prior has a smaller area than LC SAC on 2D tracking and a larger one on 3D tracking. Table~\ref{tab:bounds} examines the failures on 2D tracking, whose reference touches the upper bound $z=2$\,m. Every failure of LC SAC is the same test start, 1.3\,cm below the upper bound, failed once on each of the ten seeds early in the episode (median exit step 15). Residual SAC, the action prior and the plain threshold also fail this start on all ten seeds, and fail a few other starts as well. The failures of SAC instead come from two seeds that leave the upper bound near the end of most episodes, mostly from starts far below it, and in completed episodes SAC comes closer to the bound than the learners that use the model.

\begin{table}[H]
\centering
\caption{Learning curve area at the nominal box: mean validation return over the eight checkpoints (mean over seeds with a 95\% bootstrap interval), and after the semicolon the difference LC SAC minus the method with its interval. Ten seeds unless the count is given in parentheses. An n/a entry marks a run we did not perform.}
\label{tab:auc}
\resizebox{\textwidth}{!}{
\begin{tabular}{lllll}
\toprule
Method & 2D stab. & 2D track. & 3D stab. & 3D track. \\
\midrule
SAC & 113 [107, 119]; +31 [+25, +38] & 85 [81, 89]; +65 [+60, +71] & 82 [71, 91]; +84 [+75, +95] & 42 [38, 45]; +109 [+104, +113] \\
Shaped SAC & 104 [98, 110]; +41 [+34, +47] & 97 [93, 101]; +53 [+48, +59] & 102 [93, 111]; +64 [+55, +74] & 60 [54, 65]; +90 [+84, +97] \\
Lagrangian SAC & 118 [114, 122]; +26 [+21, +31] & 47 [40, 55]; +103 [+94, +111] & 113 [106, 120]; +53 [+46, +60] & 88 [82, 95]; +62 [+55, +69] \\
Domain rand. SAC & 118 [114, 121]; +27 [+22, +31] & 85 [83, 88]; +64 [+60, +70] & 92 [86, 97]; +75 [+69, +81] & 43 [39, 47]; +107 [+103, +112] \\
Residual SAC & 178 [176, 179]; -33 [-36, -30] & 94 [91, 98]; +56 [+50, +62] & 194 [193, 195]; -28 [-30, -25] & 181 [177, 186]; -31 [-36, -26] \\
Action prior ($w=1$) & n/a & 106 [103, 109]; +44 [+39, +49] & n/a & 170 [169, 170]; -19 [-22, -17] (5) \\
LC SAC, plain threshold & 145 [143, 148]; -1 [-4, +3] & 102 [95, 107]; +48 [+41, +56] & 165 [162, 169]; +1 [-3, +6] & 152 [147, 156]; -2 [-7, +4] \\
LC SAC & 145 [142, 147] & 150 [147, 155] & 167 [164, 169] & 150 [148, 153] \\
\bottomrule
\end{tabular}
}
\end{table}

\begin{table}[H]
\centering
\caption{2D tracking at the nominal box, test episodes of the selected checkpoint (twelve per seed, ten seeds): failed episodes, those that leave through the upper bound $z=2$\,m, the median clearance to that bound at the start of the failed episodes and their median exit step (of 250), and the median over episodes of the smallest clearance, over all and over completed episodes.}
\label{tab:bounds}
\resizebox{\textwidth}{!}{
\begin{tabular}{lrrrrrr}
\toprule
Method & failed & through top & start clearance (cm) & exit step & min.\ clearance (cm), all & min.\ clearance (cm), completed \\
\midrule
SAC & 23/120 & 23/23 & 42.0 & 248.0 & 15.6 & 20.4 \\
Residual SAC & 15/120 & 14/15 & 1.3 & 14.0 & 27.3 & 30.4 \\
Action prior ($w=1$) & 14/120 & 10/14 & 1.3 & 11.5 & 30.4 & 31.7 \\
LC SAC, plain threshold & 14/120 & 11/14 & 1.3 & 11.0 & 20.2 & 24.9 \\
LC SAC & 10/120 & 10/10 & 1.3 & 15.0 & 27.8 & 30.3 \\
\bottomrule
\end{tabular}
}
\end{table}

\subsection{Perturbations}
\label{app:robstab}

\begin{figure}[ht]
\centering
\includegraphics[width=\textwidth]{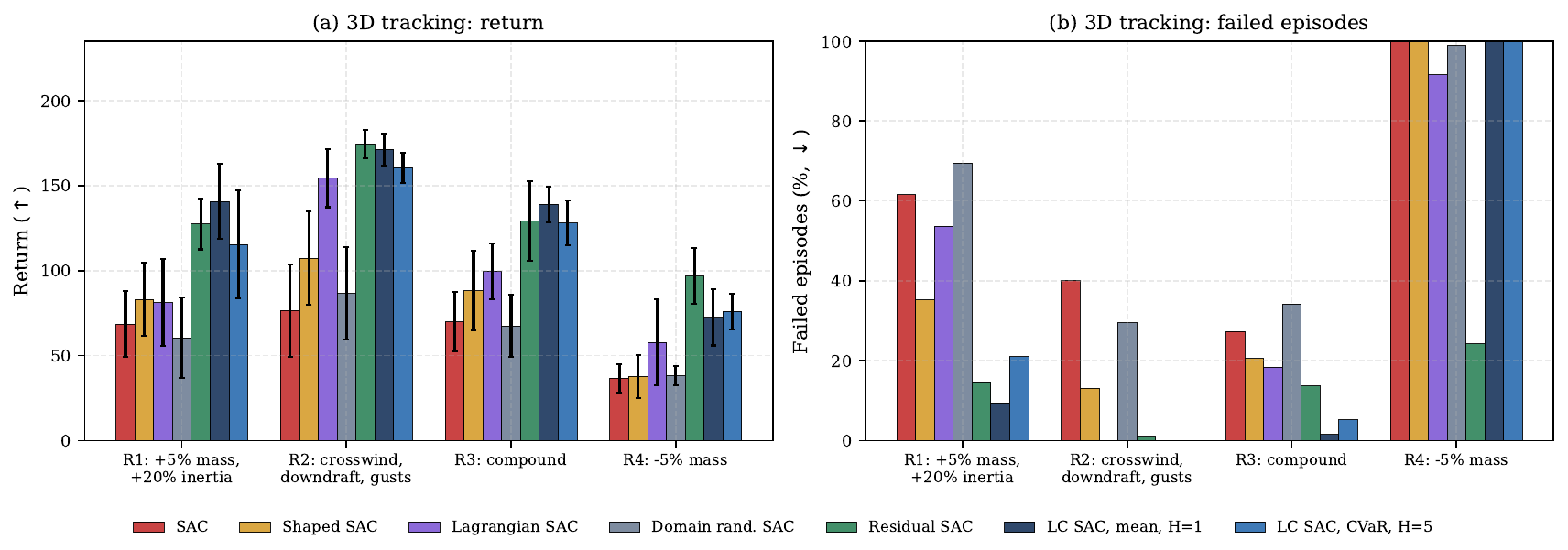}
\caption{3D tracking under the perturbations R1 to R4: return (mean $\pm$ SD over seeds, left) and failed episodes (right) on the 19 certification episodes, for the learners of Table~\ref{tab:robust}.}
\label{fig:ood}
\end{figure}

Table~\ref{tab:robust} evaluates the same checkpoints on the 19 certification episodes, without perturbation and under R1 to R4. Except for the Lagrangian SAC, whose policies mostly leave the lateral bound, nearly every nominal failure on 2D tracking is an exit through the upper bound $z=2$\,m, which the reference touches. The perturbations that push the vehicle down remove these failures: under R1 to R3 the headline variant fails in \robRoneTwoTrackHeadFail, \robRtwoTwoTrackHeadFail{} and \robRthreeTwoTrackHeadFail\% of episodes, and SAC and domain randomized SAC do not fail, while residual SAC fails in up to \robRtwoTwoTrackResFail\% (\robRoneTwoTrackResQpFail\% with the QP base) and the box QP regulator in up to \robRoneTwoTrackQpFail\%. The return of the headline variant is significantly above SAC only under R2, that of the CVaR variant with $H=5$ under all three. R4, the lighter vehicle, reverses the picture: every learner leaves the upper bound often, the headline variant in \robRfourTwoTrackHeadFail\% of episodes and the CVaR variant in \robRfourTwoTrackCvarFiveFail\%, against \robRfourTwoTrackSacFail\% for SAC, \robRfourTwoTrackResFail\% for residual SAC and \robRfourTwoTrackQpFail\% for the box QP regulator. On 3D tracking, R1 to R3 make SAC fail in up to \robRoneThreeTrackSacFail\% of episodes and the two LC SAC variants of Table~\ref{tab:robust} in up to \robRoneThreeTrackLcFailMax\%, while the fixed regulator does not fail and has the highest return under R2 and R3. Under R4, the learners fail in nearly every episode except residual SAC (\robRfourThreeTrackResFail\%), and the regulator fails in \robRfourThreeTrackQpFail\%. No learner avoids failures under perturbations in both directions, and the constraint does not help with this (Figure~\ref{fig:ood}). Table~\ref{tab:robust_stab} gives the same evaluation for the stabilization tasks.

\begin{table}[H]
\centering
\caption{Return (mean $\pm$ SD over seeds) and failure rate in \% in brackets on the 19 certification episodes of the tracking tasks, without perturbation (nominal) and under perturbations the learners never saw. Results use ten seeds except five for CVaR with $H{=}5$. $^\dagger$ (R1 to R4): better than SAC after Holm correction within the task and perturbation, even where both fail often.}
\label{tab:robust}
\resizebox{\textwidth}{!}{
\begin{tabular}{llrrrrr}
\toprule
Task & Method & Nominal & R1 & R2 & R3 & R4 \\
\midrule
2D track. & SAC & 131\,$\pm$\,8 [19] & 131\,$\pm$\,12 [0] & 126\,$\pm$\,9 [0] & 126\,$\pm$\,11 [0] & 135\,$\pm$\,12 [83] \\
 & Shaped SAC & 137\,$\pm$\,9 [13] & 139\,$\pm$\,6 [0] & 135\,$\pm$\,8 [10] & 134\,$\pm$\,6 [0] & 139\,$\pm$\,13 [73] \\
 & Lagrangian SAC & 70\,$\pm$\,31 [69] & 63\,$\pm$\,30 [61] & 70\,$\pm$\,34 [59] & 64\,$\pm$\,32 [53] & 62\,$\pm$\,28 [91] \\
 & Domain rand. SAC & 131\,$\pm$\,7 [18] & 129\,$\pm$\,9 [0] & 126\,$\pm$\,8 [0] & 124\,$\pm$\,8 [0] & 130\,$\pm$\,7 [74] \\
 & Residual SAC & 163\,$\pm$\,9 [19] & 134\,$\pm$\,25 [6] & 128\,$\pm$\,33 [18] & 126\,$\pm$\,29 [9] & 128\,$\pm$\,7 [43] \\
 & Residual SAC, QP base & 166\,$\pm$\,16 [20] & 120\,$\pm$\,46 [26] & 155\,$\pm$\,29 [14] & 129\,$\pm$\,32 [14] & 132\,$\pm$\,10 [42] \\
 & LC SAC, mean, $H{=}1$ & 172\,$\pm$\,8 [15] & 139\,$\pm$\,13 [1] & 163\,$\pm$\,11 [0]$^\dagger$ & 135\,$\pm$\,8 [0] & 95\,$\pm$\,22 [64] \\
 & LC SAC, CVaR, $H{=}5$ & 197\,$\pm$\,6 [4] & 154\,$\pm$\,3 [0]$^\dagger$ & 175\,$\pm$\,4 [0]$^\dagger$ & 149\,$\pm$\,3 [0]$^\dagger$ & 85\,$\pm$\,8 [100] \\
 & EDMD regulator, clipped & 121 [37] & 49 [79] & 92 [58] & 52 [84] & 128 [42] \\
 & EDMD regulator, box QP & 163 [21] & 84 [32] & 138 [26] & 82 [32] & 132 [42] \\
\midrule
3D track. & SAC & 84\,$\pm$\,20 [54] & 68\,$\pm$\,19 [62] & 77\,$\pm$\,27 [40] & 70\,$\pm$\,17 [27] & 37\,$\pm$\,8 [100] \\
 & Shaped SAC & 112\,$\pm$\,25 [16] & 83\,$\pm$\,22 [35] & 107\,$\pm$\,27 [13]$^\dagger$ & 88\,$\pm$\,23 [21] & 38\,$\pm$\,13 [100] \\
 & Lagrangian SAC & 176\,$\pm$\,18 [1] & 81\,$\pm$\,26 [54] & 154\,$\pm$\,17 [0]$^\dagger$ & 100\,$\pm$\,17 [18]$^\dagger$ & 58\,$\pm$\,25 [92] \\
 & Domain rand. SAC & 87\,$\pm$\,25 [47] & 61\,$\pm$\,24 [69] & 87\,$\pm$\,27 [29] & 68\,$\pm$\,18 [34] & 38\,$\pm$\,6 [99] \\
 & Residual SAC & 184\,$\pm$\,5 [4] & 127\,$\pm$\,15 [15]$^\dagger$ & 174\,$\pm$\,8 [1]$^\dagger$ & 129\,$\pm$\,23 [14]$^\dagger$ & 97\,$\pm$\,16 [24]$^\dagger$ \\
 & LC SAC, mean, $H{=}1$ & 190\,$\pm$\,6 [0] & 141\,$\pm$\,22 [9]$^\dagger$ & 171\,$\pm$\,10 [0]$^\dagger$ & 139\,$\pm$\,10 [2]$^\dagger$ & 73\,$\pm$\,17 [100]$^\dagger$ \\
 & LC SAC, CVaR, $H{=}5$ & 169\,$\pm$\,16 [1] & 115\,$\pm$\,32 [21] & 160\,$\pm$\,9 [0]$^\dagger$ & 128\,$\pm$\,13 [5]$^\dagger$ & 76\,$\pm$\,10 [100]$^\dagger$ \\
 & EDMD regulator, clipped & 179 [5] & 140 [0] & 185 [0] & 164 [0] & 82 [32] \\
 & EDMD regulator, box QP & 179 [5] & 142 [0] & 185 [0] & 165 [0] & 83 [32] \\
\bottomrule
\end{tabular}
}
\end{table}

\begin{table}[H]
\centering
\caption{Return and failure rate under perturbations on the stabilization tasks, as in Table~\ref{tab:robust}.}
\label{tab:robust_stab}
\resizebox{\textwidth}{!}{
\begin{tabular}{llrrrrr}
\toprule
Task & Method & Nominal & R1 & R2 & R3 & R4 \\
\midrule
2D stab. & SAC & 173\,$\pm$\,12 [0] & 112\,$\pm$\,32 [10] & 162\,$\pm$\,20 [0] & 89\,$\pm$\,23 [17] & 70\,$\pm$\,25 [57] \\
 & Shaped SAC & 176\,$\pm$\,8 [0] & 85\,$\pm$\,35 [5] & 127\,$\pm$\,40 [2] & 73\,$\pm$\,29 [4] & 59\,$\pm$\,33 [71] \\
 & Lagrangian SAC & 181\,$\pm$\,4 [0] & 125\,$\pm$\,34 [0] & 167\,$\pm$\,22 [0] & 101\,$\pm$\,29 [0] & 95\,$\pm$\,21 [25] \\
 & Domain rand. SAC & 176\,$\pm$\,7 [0] & 169\,$\pm$\,13 [0]$^\dagger$ & 176\,$\pm$\,6 [0] & 147\,$\pm$\,19 [0]$^\dagger$ & 112\,$\pm$\,18 [7]$^\dagger$ \\
 & Residual SAC & 193\,$\pm$\,1 [0] & 173\,$\pm$\,13 [0]$^\dagger$ & 185\,$\pm$\,3 [0]$^\dagger$ & 166\,$\pm$\,12 [0]$^\dagger$ & 148\,$\pm$\,14 [3]$^\dagger$ \\
 & LC SAC, mean, $H{=}1$ & 187\,$\pm$\,3 [0] & 115\,$\pm$\,19 [0] & 179\,$\pm$\,7 [0] & 97\,$\pm$\,14 [0] & 124\,$\pm$\,16 [5]$^\dagger$ \\
 & LC SAC, CVaR, $H{=}5$ & 182\,$\pm$\,6 [0] & 133\,$\pm$\,26 [0] & 182\,$\pm$\,3 [0] & 108\,$\pm$\,18 [0] & 105\,$\pm$\,18 [13]$^\dagger$ \\
 & EDMD regulator, clipped & 187 [0] & 159 [0] & 181 [0] & 150 [0] & 120 [0] \\
 & EDMD regulator, box QP & 187 [0] & 159 [0] & 181 [0] & 150 [0] & 120 [0] \\
\midrule
3D stab. & SAC & 161\,$\pm$\,20 [3] & 99\,$\pm$\,18 [19] & 137\,$\pm$\,29 [5] & 94\,$\pm$\,20 [18] & 20\,$\pm$\,2 [100] \\
 & Shaped SAC & 168\,$\pm$\,20 [0] & 100\,$\pm$\,12 [12] & 135\,$\pm$\,16 [3] & 95\,$\pm$\,9 [13] & 21\,$\pm$\,3 [100] \\
 & Lagrangian SAC & 187\,$\pm$\,13 [1] & 111\,$\pm$\,13 [13] & 159\,$\pm$\,21 [0] & 106\,$\pm$\,14 [20] & 24\,$\pm$\,5 [99] \\
 & Domain rand. SAC & 177\,$\pm$\,12 [0] & 117\,$\pm$\,5 [1]$^\dagger$ & 158\,$\pm$\,10 [0] & 115\,$\pm$\,4 [0]$^\dagger$ & 22\,$\pm$\,3 [100] \\
 & Residual SAC & 202\,$\pm$\,2 [0] & 146\,$\pm$\,5 [0]$^\dagger$ & 188\,$\pm$\,5 [0]$^\dagger$ & 141\,$\pm$\,8 [0]$^\dagger$ & 94\,$\pm$\,19 [24]$^\dagger$ \\
 & LC SAC, mean, $H{=}1$ & 206\,$\pm$\,2 [0] & 122\,$\pm$\,5 [2]$^\dagger$ & 174\,$\pm$\,10 [0]$^\dagger$ & 113\,$\pm$\,8 [1] & 39\,$\pm$\,18 [98]$^\dagger$ \\
 & LC SAC, CVaR, $H{=}5$ & 201\,$\pm$\,3 [0] & 126\,$\pm$\,6 [0]$^\dagger$ & 167\,$\pm$\,17 [0] & 117\,$\pm$\,5 [0]$^\dagger$ & 29\,$\pm$\,2 [100]$^\dagger$ \\
 & EDMD regulator, clipped & 188 [0] & 159 [0] & 191 [0] & 154 [0] & 37 [100] \\
 & EDMD regulator, box QP & 188 [0] & 159 [0] & 191 [0] & 154 [0] & 37 [100] \\
\bottomrule
\end{tabular}
}
\end{table}

\subsection{Supplementary Figures}
\label{app:supp}

\begin{figure}[ht]
\centering
\includegraphics[width=\textwidth]{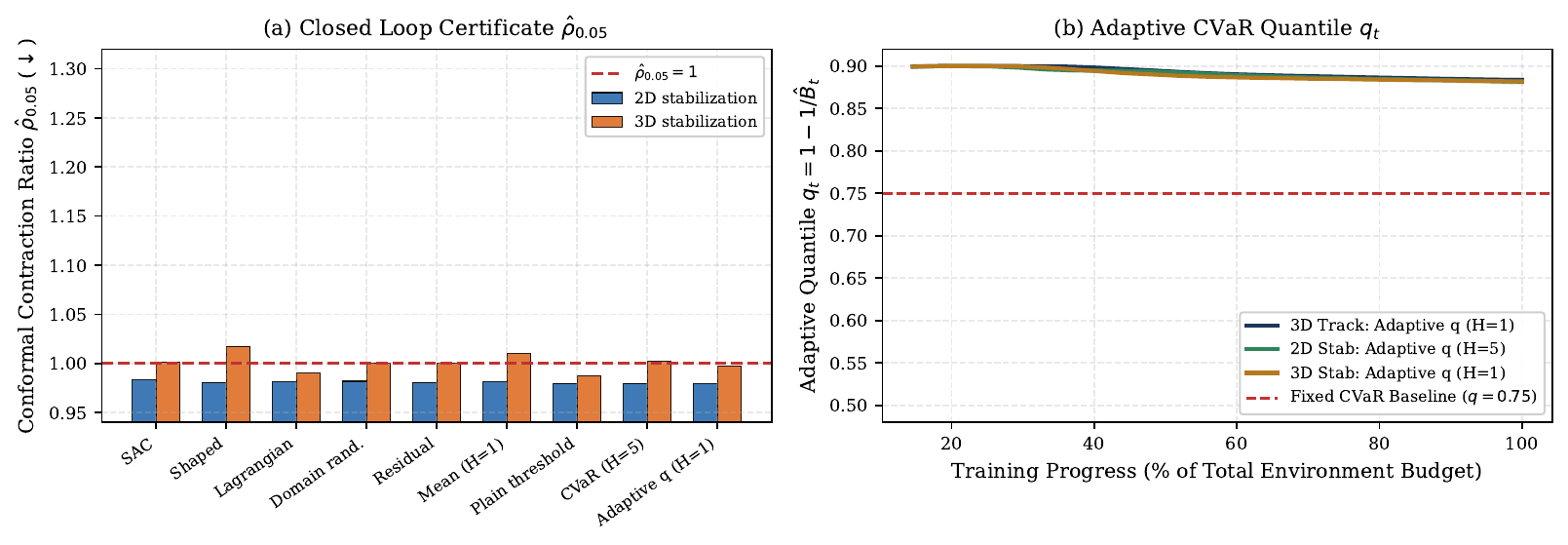}
\caption{Left: median certificate $\hat\rho$ over seeds on the stabilization tasks for the learners shown. Right: the adaptive quantile $q_t$ during training, with a mean per run between 0.88 and 0.90.}
\label{fig:q}
\end{figure}

\begin{figure}[ht]
\centering
\includegraphics[width=\textwidth]{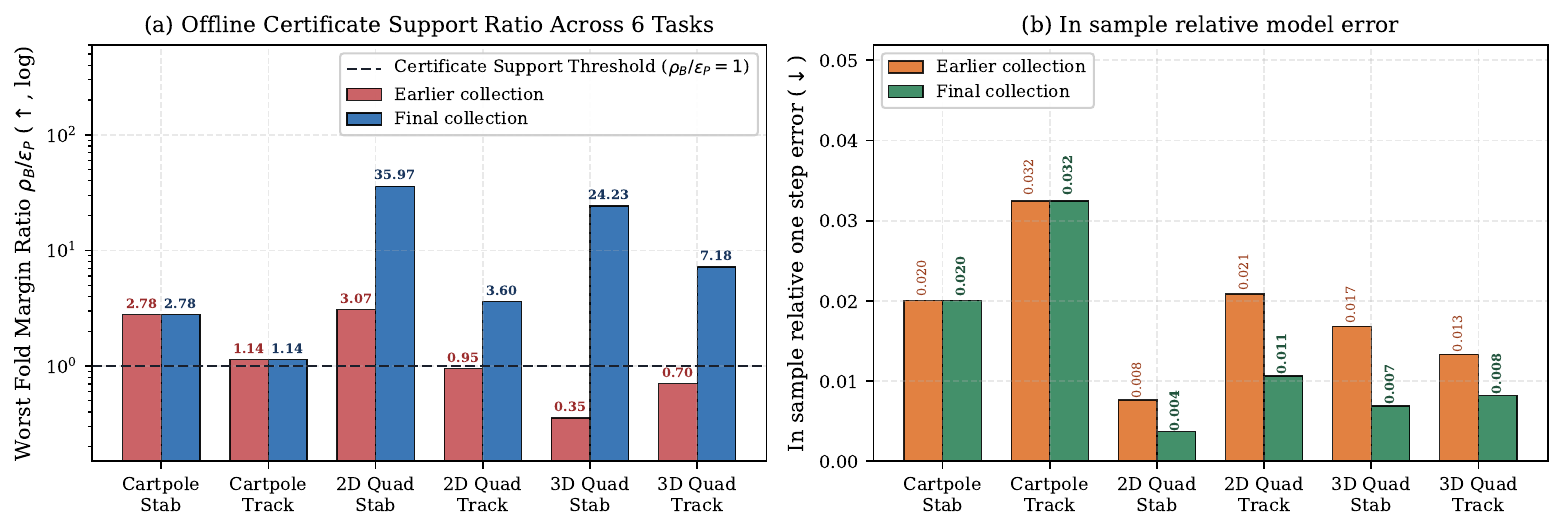}
\caption{Worst fold margin ratio $\rho_B/\varepsilon_P$ (left, log scale) and in sample relative model error (right) for the earlier and the final identification collection.}
\label{fig:certificate}
\end{figure}

\end{document}